\documentclass[a4paper,12pt]{report}
\usepackage[USenglish]{babel} %francais, polish, spanish, ...
\usepackage[T1]{fontenc}
\usepackage[ansinew]{inputenc}
\usepackage{verbatim}

\usepackage{graphicx} %%For loading graphic files
\usepackage{amsmath}
\usepackage{amsthm}
\usepackage{amsfonts}
\def\op{\ensuremath{\mathcal O}}

\usepackage{cite}
\usepackage{setspace}
\begin{document}

\pagestyle{empty} %No headings for the first pages.
%\onehalfspacing

%% Title Page %%%%%%%%%%%%%%%%%%%%%%%%%%%%%%%%%%%%%%%%%%%%%%%
%% ==> Write your text here or include other files.
%\begin{comment}
\begin{titlepage}
%\changefont{pag}{m}{n}			% AvantGarde
$$$$
\vskip 2cm
\begin{center}
\Huge \textbf{Holographic Renormalization\\for Fermions in Real Time}
\end{center}
\normalsize
\vskip 4cm

\begin{center}

\par
\begingroup
%\leftskip=4cm % ggf. verstellen
\setlength{\parindent}{0pt}  %\noindent

\vskip 1cm
\textbf{Jegors Korovins}
\vskip 1cm
LMU Munich
\vskip 1cm
A thesis submitted for the degree of \\ Master of Science\\
Munich, July 2010
\par
\endgroup

\vskip 1cm

\end{center}
\begin{tabular}{lll}
\textbf{Academic advisor}: & & Priv.-Doz. Dr. Johanna Erdmenger
\end{tabular}

\end{titlepage}
%\end{comment}
%% The simple version:
%\title{Real-time renormalization with fermions}
%\author{Jegors Korovins}
%\date{} %%If commented, the current date is used.
%\maketitle
\begin{comment}

\begin{center}

%\vspace*{1cm}
\Large
\textbf{Real-Time Holographic Renormalization \\ for Fermions}

\vspace{3cm}

%\LARGE
\textbf{Masters Thesis\\[0.5\baselineskip]
%von\\[0.5\baselineskip]
Jegors Korovins\\}
%{\normalsize \textsc{born on the 29th of February 1974 in Town}}}\\

\vspace{3cm}
\textbf{July, 2010}\\ %%Date - better you write it yourself.

\vspace{1cm}
\textbf{Supervisor:\\
Priv.-Doz. Dr. Johanna Erdmenger\\}

\vspace{1cm}
\textbf{LMU Munich\\
Max-Planck-Institute for Physics\\
(Werner-Heisenberg-Institute)}

\end{center}

\end{comment}

%% The nice version:
%\include{titlepage} %%You need a file 'titlepage.tex' for this.
%% ==> TeXnicCenter supplies a possible titlepage file
%% ==> with its templates (File | New from Template...).
\newpage
This Thesis was written at the Max Planck Institute for Physics, Munich, from February 2010
to July 2010.
\newpage

 \begin{abstract}
    We consider the real-time holography on Anti-de-Sitter (AdS) and more generally on Lifshitz spacetimes for spinorial fields. A Lifshitz spacetime has anisotropic scaling properties for the time and space coordinates. The equation of motion for fermions on general Lifshitz space is derived here for the first time. Analytically solvable cases are identified. On AdS space we derived time-ordered, time-reversed, advanced and retarded propagators with the correct $i \epsilon$-insertions. Using the Keldysh-Schwinger contour we also calculated a  propagator on thermal AdS. For massless fermions on the Lifshitz spacetime with $z=2$ we calculated the Euclidean 2-point function and explored the structure of divergences of the on-shell action for general values of $z$ and $m$. The covariant counterterm action which cancels the highest order divergence is derived.
 \end{abstract}

\newpage
\mbox{} \thispagestyle{empty}
\newpage

{						%Danksagung sichtbar
\newpage
\begin{center}
{\Large \textbf{Acknowledgements}} 
\vskip 1cm
\end{center}

\begingroup
\setlength{\parindent}{0pt} 
First of all I would like to thank my advisor Johanna Erdmenger for introducing me into the field of gauge/gravity duality and giving me an opportunity to work on my thesis in her research group. While working on the topic that she suggested, I got a feeling of the modern research.
\vskip 1cm
I am equally obliged to Andy O'Bannon. If I have learned anything about holographic renormalization, then it is mostly the outcome of numerous discussions with him.
\vskip 1cm
I am pleased to have a chance to thank Martin Ammon, whose enthusiasm made me interested in gauge/gravity duality.
\vskip 1cm
My special gratitude goes to my parents for their love, care and patience.
\vskip 1cm
Most important, I am grateful to Sveta, who made me motivated during my work on the thesis and brought new colours into the world around me.
\vskip 1cm
I am happy to thank Edgar, who brings so much fun into my life and accepts me the way I am.
\vskip 1cm
My stay in Munich would not be as productive, if not the effort of Robert Helling and Dieter L\"ust, who created a special atmosphere at the TMP programme.
\vskip 1cm
Last but not least I appreciate the support of the DAAD, that made my studies in Munich possible.

\endgroup
}

\newpage
\mbox{} \thispagestyle{empty}
\newpage

\setcounter{page}{5}

%% Inhaltsverzeichnis %%%%%%%%%%%%%%%%%%%%%%%%%%%%%%%%%%%%%%%
\tableofcontents %Table of contents
\cleardoublepage %The first chapter should start on an odd page.

\pagestyle{plain} %Now display headings: headings / fancy / ...

%% Chapters %%%%%%%%%%%%%%%%%%%%%%%%%%%%%%%%%%%%%%%%%%%%%%%%%
%% ==> Write your text here or include other files.

%\input{intro} %You need a file 'intro.tex' for this.

%%%%%%%%%%%%%%%%%%%%%%%%%%%%%%%%%%%%%%%%%%%%%%%%%%%%%%%%%%%%%
%% ==> Some hints are following:

%\chapter{Dirac equation}\label{Dirac}
\chapter{Introduction to AdS/CFT Correspondence}
AdS/CFT correspondence is one of the main achievements in theoretical physics during the last 15 years. In short, it is a conjecture saying that some conformal gauge field theory (CFT) is equivalent to the string theory on the special background called Anti de-Sitter space (AdS). The usual interpretation is that the CFT lives on the conformal boundary of AdS space. Hence AdS/CFT is a realization of holographic principle saying that for some physical systems the information in the volume (bulk) is encoded on the boundary of the volume.

For deeper understanding of this equivalence there was developed a holographic dictionary which translates physics from the one theory to another. In this thesis we address one particular piece of this dictionary: how all the different real-time correlation functions which can be computed in CFT are encoded in the bulk physics. The corresponding question for Euclidean correlators was settled already in the very first papers on the subject \cite{Maldacena:1997re, Gubser:1998bc, Witten:1998qj}, but for real-time correlators the appropriate formalism was developed only in last two years by Kostas Skenderis and Balt van Rees in a series of papers \cite{Skenderis:2008dh, Skenderis:2008dg, vanRees:2009rw}. In this thesis the emphasis is on the generalization of this formalism to the case of fermionic fields.

The structure of the thesis is as follows. The first two chapters consist of well-known material which is presented in many details in extensive literature. In the first chapter we give a short introduction into the broad subject of AdS/CFT correspondence. Second chapter is the review of the holographic renormalization for Euclidean and real-time correlators. We then generalize and apply the formalism, developed here, to the fermions in the third chapter. And finally in the fourth chapter we address the question of the renormalization of fermionic correlators on Lifshitz spacetimes. The third and fourth chapters consist of mostly new material (except sections 3.2 and 3.3).

In this chapter we will give an idea of how the AdS/CFT conjecture comes about, what is the evidence for it, etc. Details and further references may be found in a number of extensive reviews \cite{Aharony:1999ti,D'Hoker:2002aw,Horowitz:2006ct,Petersen:1999zh}.
\section{The Idea of Holography}
\subsection{D-Branes, Duality between Open and Closed Strings, Large N Expansions}
At the heart of holographic duality lies the twofold interpretation of the so-called Dp-brane solutions in supergravity (SUGRA) or superstring theory. In supergravity, Dp-branes are special solitonic (BPS) solutions of the equations of motion. Intuitively one can think of them as black hole - like ((p+1)-dimensional) objects. In particular they have a horizon. What is their meaning in superstring theory? So far there have not been obtained such solutions for the stringy equations of motion, but supergravity is well known to be the low-energy limit of superstring theory, in which the supergravity sector decouples. The important point is that such solutions are believed to be protected by supersymmetry along the renormalization group (RG) flow from the infrared (IR) to the ultraviolet (UV) region, i.e. Dp-branes should be considered as real players not only in SUGRA approximation but in the fully quantized superstring theory \cite{Green:1987sp, Green:1987mn, Polchinski:2005ST1, Polchinski:2005ST2, Johnson:2000ch}.

Another important point for the later discussion is that the low-energy limit of superstring theory is known even on the given non-trivial background - it becomes the SUGRA on that background, although it is still not known how to construct full consistent quantized superstring theory on general non-Minkowski backgrounds. Thus, the discussion above applies also in the context of AdS/CFT correspondence, where the underlying spacetime is AdS.

First interpretation of the \emph{Dp-branes in supergravity theory (and hence in superstring theory) is that they play the role of the source for the graviton}. In superstring theory gravitons are particular excitations of closed strings. They can be created by a Dp-brane, propagate and annihilate on another Dp-brane. On the other hand, string theory itself requires non-perturbative extended objects in order to incorporate open strings. Thus the second interpretation is that \emph{Dp-branes are objects which implement the Dirichlet boundary conditions for the open string.} In this case it is known that excitations of open string attached to a stack of N such objects at low energy give rise to a SU(N) gauge theory living on the volume spanned by the moving brane (so-called braneworld).

Now comes the crucial point. It is believed that these two pictures of Dp-branes in string theory can be identified (the evidence for it is based on the BPS-properties and equality of R-charges for both interpretations \cite{Johnson:2003gi}). Thus, Dp-branes play two roles in string theory, which are believed to give two equivalent interpretations of the same physical reality.

Another piece of wisdom is the discovery made by 't Hooft, that some quantum field theories may simplify, when the number of fields goes to infinity (or the rank of gauge group $N \rightarrow \infty$). The classical examples of this phenomenon include linear sigma model \cite{Peskin:1995ev} and the large $N$ matrix theories. In the latter case one can see, that in the large $N$ limit the planar diagrams give the most important contribution to the theory (one can go even further and identify these diagrams with stringy worldsheets). Surprisingly, although this situation is very different from QCD, one can still draw some important conclusions even for finite $N$ theories. For instance, if one identifies open strings with mesons one finds that they are weakly coupled and one can even reproduce the Regge trajectories.

In what follows we will often mention the $\mathcal{N} = 4$ Super-Yang-Mills theory. We do not need any detailed knowledge of it, but we say a couple of words about its properties. The $\mathcal{N} = 4$ Super-Yang-Mills theory is a gauge theory with a gauge field $A_{\mu}$, four Weyl fermions $\chi_i$, and six real scalars $\phi^I$, all in the adjoint representation of the $SU(4) \approx SO(6)$ group (group, which rotates supersymmetry generators). Its Lagrangian can be written down explicitly, but is not very important for our purposes. It has a vanishing beta function and is a scale invariant theory on quantum level (conformal group is $SO(4,2)$). The S-duality (strong - weak coupling duality) is conjectured for this theory. In practice, one works only with particular sector(s) of this theory (i.e. some subset of operators, e.g. chiral primaries) and thus the theory simplifies significantly. 

\subsection{Original Motivation}
Now we are in position to give rough idea of the original argument given by Maldacena \cite{Maldacena:1997re}, leading to the AdS/CFT conjecture. Let us consider a stack of N D3-branes in type IIB Superstring theory. The first interpretation of this situation leads to the picture in which we have usual superstring theory in the bulk, gauge theory ($\mathcal{N} = 4$ $SU(N)$ supersymmetric Yang-Mills theory) living on the brane and the interactions between these two families of fields. Schematically the resulting action is $S = S_{\text{bulk}} + S_{\text{brane}} + S_{\text{int}}$. Now we take the low energy limit. String theory provides us with the natural energy scale: we can measure energy in units of inverse string length. We keep the energy of excitations $E$ bounded and take the limit $l_s \sim \sqrt{\alpha'}\rightarrow 0$ (such that $E << 1/l_s$). \textit{ In low energy (or large distance) limit bulk action $S_{\text{bulk}}$ becomes free supergravity, $S_{\text{brane}}$ gives rise to $\mathcal{N}=4$ SU(N) Super-Yang-Mills theory which decouples from the bulk dynamics ($S_{\text{int}} \rightarrow 0$)}.

But we have also the second interpretation of this situation. D3-branes can be viewed as the sources for the gravitational field (because they have generically non-vanishing tension). Taking again low energy limit we realize that there are two kinds of massless excitations: massless fields in the bulk and fields living on the horizon of D3-branes (they appear to be massless for a distant observer because of the redshift). Again, they decouple in this limit. \textit{As a result we have two decoupled theories: free supergravity in the bulk and Type IIB superstring theory living on the near-horizon geometry of D3-branes (which happens to be $AdS_5 \times S^5$).}

Comparing these two situations, which are supposed to be equivalent, and identifying the dynamical parts one is lead to the conjecture: \textit{$\mathcal{N}=4$ SU(N) Super-Yang-Mills theory on Minkowski space is dual to the type IIB superstring theory on $AdS_5 \times S^5$.} 

This statement is extremely non-trivial. But there is some additional evidence, that these two apparently very different theories have something in common. To realize it let us think about the symmetries of these two theories. $\mathcal{N}=4$ SU(N) Super-Yang-Mills theory is well known to be a conformal field theory. Thus, it has $SO(4,2)$ conformal symmetry group. But it is exactly the isometry group of AdS$_5$! More then this, $\mathcal{N}=4$ SU(N) Super-Yang-Mills theory has global $SU(4)$ symmetry which rotates 4 SUSY generators. Again, $SU(4)$ is locally isomorphic to $SO(6)$, which is the isometry group of $S^5$! Thus, these two theories have the same global symmetries.

Type IIB superstring theory on $AdS_5 \times S^5$ is still quite complicated theory to work with. In order to simplify it one usually considers tree level supergravity theory, i.e. string coupling is sent to zero $g_s \rightarrow 0$. In addition one sends $N \rightarrow \infty$, such that $g_s N \rightarrow \infty$. On the CFT side it corresponds to the t'Hooft (or planar) limit $\lambda = g^2_{YM} N \rightarrow \infty$ and $N \rightarrow \infty$, where $\lambda$ is the effective coupling constant in field theory. This point makes the Maldacena conjecture particularly exciting. If this kind of duality is correct, then we have an access to the strong coupling limit of a \textit{quantum} field theory. Interestingly, this QFT is dual to the \textit{classical} gravity theory. On the other hand, it makes it particularly difficult to test this conjecture, since there is only very limited amount of calculations we can do in the strong coupling limit in quantum field theory.
\subsection{Geometry of AdS Spaces}
Anti-de Sitter space (AdS) appears as a (part of) geometry near the horizon of D-branes. Therefore we review here some of the most important properties of AdS space. AdS$_{d+1}$ space is a homogeneous (i.e. each point can be transformed into another one by an isometry) isotropic space with constant negative curvature. It can be embedded in $\mathbf{R}^{d+2}$ as a hyperboloid
\begin{equation}
X^2_0 + X^2_{d+1} - \sum^{d}_{i=1} X^2_i = L^2,
\label{eq:metr}
\end{equation}
with the metric
\begin{equation}
ds^2 = -dX^2_0 - dX^2_{d+1} + \sum^{d}_{i=1} dX^2_i.
%\label{eq:metr}
\end{equation}
$L$ is called the radius of AdS. (\ref{eq:metr}) can be solved by setting
\begin{align}
X_0 = L \cosh \rho \cos \tau, \qquad X_{d+1} = L \cosh \rho \sin \tau \nonumber \\ 
X_i = L \sinh \rho \; \Omega_i (i = 1,...,d; \sum_i \Omega^2_i=1),
\label{global}
\end{align}
with $\rho \geq 0$ and $0\leq \tau \leq 2 \pi$. These coordinates cover the hyperboloid once and are called ''global coordinates''. The topology of this hyperboloid is $S^1 \times \mathbf{R}^{d}$, with $S^1$ representing closed timelike curves in the $\tau$ direction. To obtain a causal spacetime we simply unwrap this circle and obtain the universal covering of the original hyperboloid with no closed timelike curves.

We list some important properties of AdS$_{d+1}$:
\begin{itemize}
\item The isometry group is $SO(2,d)$.
\item It has a $d$-dimensional conformal boundary.
\item The cosmological constant is negative, $0 > \lambda = -\frac{1}{L^2} d (d-1) $.
\item Massive fields can never get to the conformal boundary, but massless fields can go to the boundary and back in finite proper time.
\item Field theories involving negative mass fields can still be stable (there is the so-called Breitenlohner - Freedman bound on the mass of scalar field: $m^2 \geq -\frac{d^2}{4}$).
\end{itemize}

There are many kinds of coordinates for $AdS$ spaces. In addition to the 'global' parameterization (\ref{global}) there is another set of coordinates often used in the literature ($y, t, \vec{x}$) ($0<y, \vec{x} \in \mathbf{R}^{d-1}$). It is defined by
\begin{align}
X_0 &= \frac{1}{2y} (1 + y^2 (1 + \vec{x}^2 - t^2)), \nonumber \\
X_i &= y x_i, \nonumber \\
X_d &= \frac{1}{2 y} (1 - y^2 (1 - \vec{x}^2 + t^2)), \nonumber \\
X_{d+1} &= yt,
\end{align}
where we set the radius of AdS $L=1$. These coordinates cover one half of the hyperboloid (\ref{eq:metr}). In new coordinate the metric becomes
\begin{align}
ds^2 = \left ( \frac{dy^2}{y^2} + y^2(-dt^2 + d\vec{x}^2) \right).
\end{align}

For the discussion of renormalization, the so-called Fefferman - Graham coordinates are very convenient. One obtains them by setting $u=1/y$. In those the metric takes the form
\begin{equation}
ds^2 = \frac{du^2 + \eta_{ij}dx^i dx^j}{u^2}.
\label{FG}
\end{equation}
The $u\geq 0$ coordinate represents the radial direction and the conformal boundary is at $u = 0$. In this form one also sees explicitly that this metric is scale invariant (invariant with respect to the scalings $x \rightarrow \lambda x, u \rightarrow \lambda u$).

Often it is enough if the spacetime is only asymptotically AdS (AAdS), i.e. it is approaching AdS geometry near the conformal boundary. In this case we can replace $\eta_{ij}$ in the above expression by some more general metric $g_{ij} (x)$ which is approaching $\eta_{ij}$ when $u \rightarrow 0$.
\subsection{Prescription} 
Now it is time to clarify in which sense these theories are equivalent. It is important to observe that $\mathcal{N}=4$ SU(N) Super-Yang-Mills theory, being a conformal theory, does not possess an S-matrix, i.e. one can only speak about the correlation function of gauge-invariant operators (gravity cannot have any clue about the $SU(N)$ gauge symmetry). The basic idea is to identify the generating functional of connected correlators in the superconformal gauge theory with the minimum of the supergravity action, subject to some boundary conditions. To be more concrete, think of scalar field $\phi$ of the mass $m$ in the bulk. Let $O$ be its dual operator of conformal dimension $\Delta$ (which is related to the mass $m$) on the field theory side (i.e. $O$ lies in the same representation of global symmetries as $\phi$). There are two linearly independent solutions to the equation of motion for $\phi$ which are characterized by their boundary behavior. One mode is normalizable and another is non-normalizable. Non-normalizable modes have some given boundary behavior $\phi \rightarrow u^{d-\Delta} \phi_0$. We identify this $\phi_0$ with the source for $O$. The basic prescription then says that the supergravity partition function (which is a functional of the fields parameterizing the boundary behavior of the bulk fields) is identified with the generating functional of QFT correlation functions
\begin{equation}
\left\langle \exp[\int \!{d^d x \phi_0 O}] \right\rangle_{CFT} = e^{-S_{\text{on-shell}}[\phi_0]},
\label{eq:prescription}
\end{equation}
where $S_{\text{on-shell}} [\phi_0]$ is the supergravity action evaluated on the \textit{regular} solution with the given asymptotic behavior and is viewed as a functional of the boundary value $\phi_0$. This prescription fails in Lorentzian signature, since in that case generically there is no regular solution to the equation of motion.

Let us summarize basic points of AdS/CFT prescription:
\begin{itemize}
\item The background solution is associated with the vacuum of the dual QFT. Perturbations around the background are associated with correlation functions of gauge invariant operators.

\item The isometries of the bulk solution correspond to global symmetries of the boundary theory. Recall that the AdS group in $d+1$ dimensions $SO(d,2)$ coincides with the conformal group in $d$ dimensions.

\item Gauge invariant operators of the boundary theory are in one-to-one correspondence with bulk fields. For example, the bulk metric corresponds to the stress energy tensor of the boundary theory.

\item In a spacetime with a boundary one needs to specify boundary conditions for the bulk fields. The leading boundary behavior of the bulk field is identified with the source $\phi_0$ of the dual operator.

\item Correlation function can now be computed by functionally differentiating with respect to the sources. For example,
\begin{align}
\left\langle O(x)\right\rangle &= \frac{\delta S_{\text{on-shell}}}{\delta \phi_0(x)} \\
\left\langle O(x_1) O(x_2)\right\rangle &= - \frac{\delta^2 S_{\text{on-shell}}}{\delta \phi_0(x_1) \delta \phi_0(x_2)}
%\label{eq:}
\end{align}

\item A naive use of these formulas however yields infinite answers. The on-shell value of the action is infinite due to the infinite volume of the AAdS spacetime. Similarly, the QFT correlators diverge and need to be renormalized. The goal of holographic renormalization is to make such formulae well-defined.
\end{itemize}

\section{Some Tests and Extensions of AdS/CFT Correspondence}
Equation (\ref{eq:prescription}) is the basic prescription of AdS/CFT correspondence. So far we gave only heuristic arguments for it. But there are many calculations which can be done in order to test this duality. One can calculate 2-point \cite{Gubser:1998bc} and 3-point functions \cite{Freedman:1998tz}, match the spectra of two theories, calculate conformal anomalies \cite{Henningson:1998gx}. The interested reader is invited to consult extensive literature on this subject. Here we want to note that calculations in strongly coupled $\mathcal{N}=4$ $SU(N)$ Super-Yang-Mills theory can generically be done only for the quantities which are protected by the supersymmetric non-renormalization theorems. Then one can perform perturbative calculations in the weak coupling limit. So far there was found no mismatch when these calculations were compared with gravitational ones. Gravity produces always the correct results for the quantities we can calculate on CFT side.

The numerous tests of the AdS/CFT correspondence made people believe that this duality should hold also in some other situations. Several extensions of AdS/CFT correspondence proved to be very plausible and useful. In different settings conformal symmetry or some amount of supersymmetry is broken. In this context one talks generally about gauge/gravity duality. For example one can introduce the finite temperature by putting a black hole in the bulk \cite{Witten:1998zw}. The Hawking temperature of the black hole corresponds to the temperature on the field theory side. Giving a charge to the black hole results in introducing chemical potential in the field theory. Using standard techniques such as lattice gauge theory, it is so far nearly impossible to get any numerical results for dynamical processes in strongly coupled systems with finite temperature and chemical potential. Hence gauge/gravity is so far the only source of reliable results for such systems. One of the most famous results is the bound on the shear viscosity - entropy density ratio: $\frac{\eta}{s} \geq \frac{1}{4 \pi}$ \cite{Policastro:2002se, Son:2007vk}. Holographic realizations of further effects (e.g. different phase transitions, chiral symmetry breaking \cite{Babington:2003vm}, superfluidity and superconductors \cite{Hartnoll:2008vx, Ammon:2008fc, Ammon:2009fe}, etc.) were also found.

In recent years there was a lot of work devoted to the fermions in gauge / gravity duality. It was found, that holographic models open a window to the understanding of many interesting phenomena in strongly coupled condensed matter physics, such as superconductivity, superfluidity, quantum criticality, etc \cite{Liu:2009dm, Faulkner:2010da}. On the search of holographic dual to quantum chromodynamics (the so-called AdS/QCD correspondence) fermions also play an important role. Clearly fermionic fields deserve the attention we pay to them in this thesis.

\chapter{Holographic Renormalization}
In this chapter we give a brief review of the formalism of holographic renormalization. Simultaneously we provide the precise recipe, how to calculate renormalized QFT correlation functions using the physics in the bulk. For the case of Euclidean signature we follow mostly the pedagogical introduction of \cite{Skenderis:2002wp}. The appropriate formalism for real-time renormalization was introduced in \cite{Skenderis:2008dh, Skenderis:2008dg}.
\section{Euclidean Signature}
\subsection{Basic Idea and Example(s)}
As already mentioned the prescription (\ref{eq:prescription}) is only a formal equality. Generically both sides of it are infinite. To cure this obstacle we must subtract the infinities adding covariant counterterms. The short recipe is provided in \cite{Skenderis:2002wp}:
\begin{enumerate}
\item Compute the most general asymptotic solution of the bulk field equations.
\item To regulate the divergences we restrict the radial coordinate to have a finite range $u\geq \epsilon$, and evaluate the boundary term at $u=\epsilon$ on the regular solution:
\begin{equation}
S_{\text{reg}}[\phi, \epsilon] = S_{\text{on-shell}}[\phi(u = \epsilon)]
%\label{eq:}
\end{equation}
\item We evaluate the action on the asymptotic solutions and isolate the terms which diverge as $\epsilon \rightarrow 0$.
\item We subtract the infinite terms by adding suitable covariant counterterms $S_{CT}$:
\begin{equation}
S_{CT}[\phi(x,u = \epsilon)] = - \text{divergent terms of} \, S_{\text{reg}}[\phi, \epsilon],
%\label{eq:}
\end{equation}
where $S_{CT}$ must be expressed in terms of the fields living on regulating surface $u = \epsilon$ and the induced metric $\gamma_{ij} = g_{ij}(x,\epsilon) / \epsilon$. This is needed to ensure the covariance.

\item We define a subtracted action at the cutoff
\begin{equation}
S_{\text{sub}}[\phi(x,u=\epsilon)] = S_{\text{reg}}[\phi, \epsilon] + S_{CT}[\phi(x,u = \epsilon)].
%\label{eq:}
\end{equation}
It has a finite limit as $\epsilon \rightarrow 0$.

\item The renormalized action is then given by
\begin{equation}
S_{\text{ren}}[\phi_0] = \lim_{\epsilon \rightarrow 0} \left(S_{\text{reg}}[\phi, \epsilon] + S_{CT}[\phi(x,u = \epsilon)]\right)
\label{Sren}
\end{equation}
We need to distinguish between $S_{\text{sub}}$ and $S_{\text{ren}}$ because the variations required to obtain correlation functions are performed before the limit $\epsilon \rightarrow 0$ is taken.

\item Exact $1$-point function is obtained by differentiating the subtracted action with respect to the field on the regulating boundary and then taking the limit $\epsilon \rightarrow 0$:
\begin{equation}
\left\langle O(x) \right\rangle = \lim_{\epsilon \rightarrow 0} \left( \frac{1}{\epsilon^{\frac{d}{2} -m}} \frac{1}{\sqrt{\gamma}}\frac{\delta S_{\text{sub}}}{\delta \phi(x , u = \epsilon)}\right),
%\label{eq:}
\end{equation}
where $\gamma$ is the determinant of the induced metric and $m$ is the divergence degree of the source.

\item From the renormalized $1$-point function all the other renormalized $n$-point functions containing the same operator $O$ can be obtained by the differentiation with respect to the source $\phi_0$.

\end{enumerate}

For completeness we would like to mention, that there is yet another technique of holographic renormalization which is based on Hamiltonian formulation and is extremely useful for practical calculations \cite{Papadimitriou:2004ap, Papadimitriou:2004rz}.

\subsection{Example}
Next, we want to illustrate this recipe on a simple example: Massive scalar on pure AdS (see \cite{Skenderis:2002wp} for more details). We take the metric of AdS in the form
\begin{equation}
ds^2 = \frac{d\rho^2}{4 \rho^2} + \frac{1}{\rho} dx^i dx^i,
%\label{eq:}
\end{equation}
where we put $\rho = u^2$ in (\ref{FG}).
The action for the massive scalar field $\Phi$ is
\begin{equation}
S = \frac{1}{2}\int \!{d^{d+1} x \sqrt{G} (G_{\mu \nu} \partial_{\mu} \Phi \partial_{\nu} \Phi + m^2 \Phi^2)}.
%\label{eq:}
\end{equation}
The equation of motion is
\begin{equation}
(- \Box_G + m^2) \Phi = -\frac{1}{\sqrt{G}} \partial_{\mu} (\sqrt{G} G^{\mu \nu} \partial_{\nu} \Phi) + m^2 \Phi =0.
%\label{eq:}
\end{equation}
This equation can be solved analytically on pure AdS but we first outline the holographic procedure on asymptotically AdS (AAdS) spacetime.

First, we write the asymptotic expansion for a solution. The equation of motion is second order, hence we look for a solution of the form
\begin{align}
\Phi(x,\rho) &= \rho^{\frac{d-\Delta}{2}}\phi(x,\rho) \nonumber \\ &=\rho^{\frac{d-\Delta}{2}} (\phi_{(0)}(x) + \phi_{(0)}(x) \rho + ... + \rho^n (\phi_{(2 n)}(x) + \ln \rho \psi_{(2n)}(x))+...),
%\label{eq:}
\end{align}
where $\phi_{(0)}(x)$ corresponds to a source (boundary condition), $\phi_{(2 n)}(x)$ - to a responce, and $\psi_{(2n)}(x)$ - to the matter conformal anomaly \cite{Henningson:1998gx}. Setting this solution back into the equation of motion we get
\begin{equation}
(m^2 - \Delta (\Delta - d)) \phi - \rho (\Box_0 \phi + 2(d + 2 - 2 \Delta) \partial_{\rho} \phi + 4 \rho \partial^2_{\rho} \phi) = 0,
\label{eq:1}
\end{equation}
where $\Box_0 = \delta^{ij} \partial_i \partial_j$ is the D'Alambertian on the boundary. The easiest way to solve (\ref{eq:1}) is to successively differentiate with respect to $\rho$ and then set $\rho = 0$. In this way we obtain
\begin{equation}
m^2 = \Delta (\Delta - d),
%\label{eq:}
\end{equation}
which is the well-known relation between the mass of the scalar field in the bulk and the conformal weight of the dual operator in the bulk. (\ref{eq:1}) reduces to
\begin{equation}
\Box_0 \phi + 2(d + 2 - 2 \Delta) \partial_{\rho} \phi + 4 \rho \partial^2_{\rho} \phi = 0.
\label{eq:2}
\end{equation}
Setting $\rho = 0$ we get an algebraic equation for $\phi_{(2)}$, which is solved by
\begin{equation}
\phi_{(2)} = \frac{1}{2 (2\Delta -d -2)} \Box_0 \phi_{(0)}.
%\label{eq:}
\end{equation}
Differentiate (\ref{eq:2}) with respect to $\rho$ and then set $\rho = 0$. The result is
\begin{equation}
\phi_{(4)} = \frac{1}{4 (2\Delta -d -4)} \Box_0 \phi_{(2)}.
%\label{eq:}
\end{equation}
Continuing this way we can obtain almost all the coefficients $\phi_{(2 j)}$. This procedure stops, however, when $2\Delta - d - 2 n = 0$. At this order we have to introduce the logarithmic term to obtain a solution. For concreteness consider the case $2\Delta - d - 2 = 0$, i.e. $\Delta = \frac{d}{2} + 1$. The asymptotic expansion is given by
\begin{equation}
\phi(x,\rho) = \phi_{(0)} + \rho (\phi_{(2)} + \ln \rho \psi_{(2)})+...
%\label{eq:}
\end{equation}
Inserting this equation into (\ref{eq:2}) gives
\begin{equation}
\psi_{(2)} = -\frac{1}{4} \Box_0 \phi_{(0)}
%\label{eq:}
\end{equation}
and we find that $\phi_{(2)}$ is not determined by the asymptotic analysis. It can be found using the regular analytic solution to the equation of motion. 

We are now in position to evaluate the regularized action on the asymptotic solution,
\begin{align}
S_{\text{reg}} &= \frac{1}{2} \int_{\rho \geq \epsilon} \!{d^{d+1} x \sqrt{G} ( G_{\mu \nu} \partial_{\mu} \Phi \partial_{\nu} \Phi + m^2 \Phi^2 )} \nonumber \\ &= \frac{1}{2} \int_{\rho \geq \epsilon} \!{d^{d+1} x \sqrt{G} \Phi(- \Box_G + m^2) \Phi} - \frac{1}{2} \int_{\rho = \epsilon} \!{d^x \sqrt{G} G^{\rho \rho} \Phi \partial_{\rho}\Phi}. 
%\label{eq:}
\end{align}
The bulk term vanishes on the solution to the equation of motion and we can isolate the divergent terms
\begin{equation}
S_{\text{reg}} = \int_{\rho = \epsilon} \!{d^d x (\epsilon^{-\Delta + \frac{d}{2}} a_{(0)} + \epsilon^{-\Delta + \frac{d}{2} + 1} a_{(2)} + ... -\ln \epsilon a_{(2\Delta - d)})},
%\label{eq:}
\end{equation}
where the coefficients $a_{(2 i)}$ are local functions of the source $\phi_{(0)}$:
\begin{align}
a_{(0)} &= -\frac{1}{2} (d-\Delta) \phi^2_{(0)}, \; a_{(2)} = -(d-\Delta +1) \phi_{(0)}\phi_{(2)}... \nonumber \\
a_{(2\Delta -d)} &= - \frac{d}{2^{2 \Delta - d} \Gamma(2 \Delta - d) \Gamma(2 \Delta - d +1)} \phi_{(0)}(\Box_0)^{2 \Delta - d} \phi_{(0)}. 
%\label{eq:}
\end{align}

Now we want to find the covariant counterterms $S_{CT}$ which cancel the divergences in $S_{\text{reg}}$. For this we need to reexpress $\phi_{(0)}$ in terms of $\Phi(x,\epsilon)$ (for covariance). To second order we obtain
\begin{align}
\phi_{(0)} &= \epsilon^{-\frac{d-\Delta}{2}} \left( \Phi(x,\epsilon) - \frac{1}{2(2\Delta - d -2)} \Box_{\gamma} \Phi(x,\epsilon)\right), \nonumber \\
\phi_{(2)} &= \epsilon^{-\frac{d-\Delta}{2} -1} \frac{1}{2(2\Delta - d -2)} \Box_{\gamma} \Phi(x,\epsilon),
%\label{eq:}
\end{align}
where $\Box_{\gamma}$ is the Laplacian of the induced metric $\gamma_{ij} = \frac{\delta_{ij}}{\epsilon}$ at $\rho = \epsilon$. It is sufficient to rewrite $a_{(0)}$ and $a_{(2)}$ in terms of $\Phi(x,\epsilon)$. The counterterm action is then given by
\begin{equation}
S_{CT} = \int \!{d^d x \sqrt{\gamma} \left( \frac{d-\Delta}{2}\Phi^2 + \frac{1}{2(2\Delta - d-2)} \Phi \Box_{\gamma}\Phi + \text{higher derivatives}  \right )}.
%\label{eq:}
\end{equation}
Notice, that when $\Delta = d/2+1$ the coefficient of $\Phi \Box_{\gamma}\Phi$ is replaced by $-\frac{1}{4}\ln \epsilon$. Similarly, when $\Delta = d/2+k$ there is a $k$-derivative logarithmic counterterm.

$S_{\text{ren}}$ is now given by (\ref{Sren}). We still can add finite counterterms to it. This corresponds to the scheme dependence on the field theory side.

Renormalized $1$-point function is
\begin{equation}
\left\langle O_{\Phi} \right\rangle = \lim_{\epsilon \rightarrow 0} \left( \frac{1}{\epsilon^{\frac{\Delta}{2}}} \frac{1}{\sqrt{\gamma}} \frac{\delta S_{\text{sub}}}{\delta \Phi(x,\epsilon)}  \right).
%\label{eq:}
\end{equation}
For concreteness we discuss the $\Delta = \frac{d}{2} + 1$ case. Now,
\begin{align}
\delta S_{\text{sub}} &= \delta S_{\text{reg}} + \delta S_{CT} \nonumber \\
&= \int_{\rho \geq \epsilon} \!{d^{d+1} x \sqrt{G} \delta \Phi (-\Box_G +m^2)\Phi} \nonumber \\ &\hspace{2cm}+ \int_{\rho = \epsilon} \! d^d x \sqrt{\gamma} \delta \Phi \left( -2 \epsilon \partial_{\epsilon}\Phi + (d-\Delta)\Phi - \frac{1}{2} \ln \epsilon \Box_{\gamma}\Phi \right).
%\label{eq:}
\end{align}
On shell
\begin{equation}
\frac{\delta S_{\text{sub}}}{\delta \Phi} = \sqrt{\gamma} (-2 \epsilon \partial_{\epsilon}\Phi + (d-\Delta)\Phi - \frac{1}{2} \ln \epsilon \Box_{\gamma}\Phi).
%\label{eq:}
\end{equation}
Substituting for $\Phi$ the explicit asymptotic expansion we find that the divergent terms cancel, as promised, and the finite part equals
\begin{equation}
\left\langle O_{\Phi} \right\rangle = - 2 (\phi_{(2)}+\psi_{(2)}).
\label{eq:O}
\end{equation}
Here we see that, indeed, $\phi_{(2)}$ correspond to the responce to the perturbation and is not determined by the asymptotic analysis. This is very generic feature of such calculations: to fix $1$-point function we need regular solution. $\psi_{(2)}$ term is actually scheme dependent. One can remove it completely by adding to the $S_{CT}$ a finite term proportional to the conformal anomaly.

So far we investigated only near-boundary behavior. Holographic $1$-point function involves a coefficient which is not determined by asymptotic analysis. Now we solve the equation of motion analytically, impose regularity in the bulk and get this coefficient. For definiteness we work in $d=4$ and consider the case $\Delta = d/2+1 = 3$. We change radial variable $\rho = u^2$ and $\Phi = u^{d/2} \chi$, and we also Fourier transform in boundary directions (we have Euclidean signature on the boundary theory). The equation for $\chi$ is
\begin{equation}
u^2 \partial_{u}^2 \chi + u \partial_{u} \chi - (k^2 u^2 + 1)\chi = 0.
%\label{eq:}
\end{equation}
The regular solution is
\begin{equation}
\chi = K_1(k u) = \frac{1}{k u} + \left( \frac{1}{4}(-1 + 2 \gamma) + \frac{1}{2}(- \ln 2 + \ln k u)  \right) k u + ...,
%\label{eq:}
\end{equation}
where we have expanded the modified Bessel function $K$ near the boundary of the bulk $u = 0$ ($k = \left| k \right|$). Converting back to $\rho$ coordinate we get
\begin{equation}
\Phi(k,\rho) = \rho^{\frac{d-\Delta}{2}} \phi_{(0)}(k) \left( 1 + \rho \left( (\frac{1}{4}(-1 + 2\gamma) + \frac{1}{2}\ln \frac{k}{2}) k^2 + \frac{1}{4} k^2 \ln \rho \right)  \right) + ....
%\label{eq:}
\end{equation}
We now read off
\begin{align}
\psi_{(2)}(k) &= \frac{1}{4} k^2 \phi_{(0)}(k) \rightarrow \psi_{(2)}(x) = - \frac{1}{4} \Box_0 \phi_{(0)}(x), \\
\phi_{(2)}(k) &= \phi_{(0)}(k) \left( \frac{1}{4}(-1 + 2\gamma) + \frac{1}{2} \ln \frac{k}{2} \right) k^2.
%\label{eq:}
\end{align}
Notice that the exact solution correctly reproduces the value for $\psi_{(2)}$ as determined by the near boundary analysis. We found also that $\phi_{(2)}$ is related non-locally to the source $\phi_{(0)}$. That is why it is impossible to get it from the asymptotic analysis.

Inserting lst two equations back in (\ref{eq:O}) we get 
\begin{equation}
\left\langle O_{\Phi}(k) \right\rangle = - 2 \phi_{(0)}(k) \left[ \left( \frac{1}{4}(-1 + 2\gamma) -\frac{1}{2} \ln 2 + \frac{k^2}{4} \right) + \frac{k^2}{4}\ln k^2 \right].
%\label{eq:}
\end{equation}
The terms in parenthesis lead to contact terms in the $2$-point function and can be dropped out.
When the renormalized $1$-point function is known as a functional of the source all the other renormalized correlation functions can be obtained by differentiating $1$-point function with respect to the source. We get
\begin{equation}
\left\langle O_{\Phi}(k) O_{\Phi}(-k)\right\rangle = \frac{\delta \left\langle O_{\Phi}(k)\right\rangle}{\delta \phi_{(0)}(-k)} = \frac{1}{2}k^2 \ln k^2.
%\label{eq:}
\end{equation}
This is the correct form for the $2$-point function of the operator of the conformal dimension $\Delta = d/2+1 = 3$.
%\subsection{Hamiltonian Approach}
\section{Lorentzian Signature}
First calculations in AdS/CFT correspondence relied on the recipe provided in \cite{Gubser:1998bc, Witten:1998qj} for the Euclidean correlators. Working in the Euclidean signature is a common and convenient practice, since usually one can analytically continue Euclidean correlators to the case of Minkowski signature. In many cases, however, it is desirable to extract the real-time correlators directly from gravity. Many important and interesting properties of gauge theories at finite temperature and finite density, most notably the response of the thermal ensemble to small perturbations that drive it out of equilibrium, can only be learned from real-time Green's functions.

From the more theoretical point of view one would like to understand the interplay between causality and holography. Since bulk and boundary light cones differ from each other it is not a priori clear that bulk computation provide the correct causal structure. More specifically, we want to study dynamical processes (or processes on time-dependent backgrounds) such as gravitational collapse.

For some years it was a real challenge to generalize the Euclidean recipe for real-time correlators. The main difficulty is the following: in Euclidean signature the requirements of the regularity in the bulk and normalizability on the boundary determine the solution to the bulk equation of motion uniquely. When we consider the boundary Lorentzian signature, this is not the case anymore. Generically, in order to construct regular solution one must sum infinitely many normalizable solutions. A related issue is that in Lorentzian case one also has to specify initial and final conditions for the bulk fields. These conditions should be related to a choice of in- and out-state in the Lorentzian boundary of QFT.

It leads to the question: which condition one has to impose in the interior of the bulk? Already in late 1990s it was conjectured that different conditions in the bulk correspond to the manifold of different correlators one can calculate in real-time QFT \cite{Balasubramanian:1998sn, Balasubramanian:1998de, Balasubramanian:1999ri}. There is one particular choice of such a condition which looks especially natural: look for an infalling wave solution, i.e. for a solution which describes a wave moving toward the horizon. Such a choice should correspond to the time-ordered (Feynman) correlator on the field theory side. This recipe was first put forward in \cite{Son:2002sd} and since then used widely for performing real-time calculations. In spite of its power it has a couple of serious drawbacks. First, this prescription can be applied only for the calculation of $2$-point functions. Second, the existence of a horizon is assumed in the bulk. This is somewhat unsatisfactory, since from the holographic point of view all the information should be encoded only on the boundary of the spacetime.

Recently, new approach to this problem was developed in \cite{Skenderis:2008dh, Skenderis:2008dg}. The starting point there is the observation, that different real-time correlators can be specified by the choice of the contour in a complex time plane. Examples are given in the figure \ref{con}. In \cite{vanRees:2009rw} it was explained when the new construction is equivalent to the imposing infalling boundary condition at the horizon.

Taking principles of the holography seriously one should reflect the choice of the contour on the gravitational side too. The ingenious idea in \cite{Skenderis:2008dh} is to start with a QFT time contour and 'fill it in' with a bulk manifold. It is, real segments of the contour are associated with the Lorentzian spacetime, and imaginary segments - with Euclidean solutions. The Euclidean bulk solution which is associated with the initial state on the QFT side can also be thought of as providing a Hartle-Hawking wave function for the bulk theory \cite{Hartle:1983ai}.

In next subsections we discuss the real-time prescription of Skenderis and van Rees in some more detail. For a comprehensive review consult \cite{Skenderis:2008dg}.

\subsection{Real-time QFT}
\begin{figure}
\centering
\includegraphics{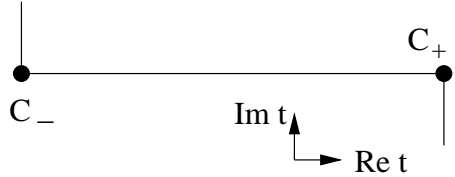}
\caption{This contour in the time plane produces the time-ordered correlator. The figure is taken from \cite{Skenderis:2008dh}.}
\label{Feynman}
\end{figure}

We shall illustrate the main idea on the example of a scalar field. Consider a field configuration with initial condition $\phi_-(\vec{x})$ at $t=-T$ and final condition $\phi_+(\vec{x})$ at $t=T$. To get the transition amplitude $\left\langle \phi_+,T|\phi_-,-T \right\rangle$ one has to integrate over all the field configurations constrained to satisfy these conditions
\begin{equation}
\left\langle \phi_+,T|\phi_-,T \right\rangle = \int_{\phi(\pm T) \! = \phi_{\pm}}{\mathcal{D}\phi e^{i S[\phi]}}
%\label{eq:}
\end{equation}
If we are interested in vacuum amplitudes we multiply this amplitude by $\left\langle 0|\phi_+,T  \right\rangle$ and $\left\langle \phi_-,T|0 \right\rangle$ and integrate over intermediate configurations $\phi_+$ and $\phi_-$. The multiplications with these vacuum wave functions correspond to extending the fields in the path integral to live on the vertical segments in the complex time plane as shown if figure \ref{Feynman}. Indeed, the infinite vertical segment starting at $-T$ corresponds to an amplitude $\lim_{\beta \rightarrow \infty} \left\langle \phi_-,-T|e^{-\beta H}|\Psi \right\rangle$ for some state $\Psi$ which is irrelevant, since the limit projects it onto the vacuum state. Similarly we obtain $\left\langle 0|\phi_+,T  \right\rangle$ from the vertical segment starting at $t = T$.

Thus, we can use the Euclidean path integral in order to create the vacuum state which is then used to constrain the Lorentzian path integral. Or, we can compute correlators in non-trivial states. Similarly, in conformal field theory there is the notion of operator - state correspondence: inserting a local operator $\mathcal{O}$ at the origin of space $\mathbf{R}^d$ and then performing the path integral over the interior of the sphere $S^{d-1}$ that surrounds the origin results in the corresponding quantum state $\Psi_{\mathcal{O}}$ on $S^{d-1}$. In particular, the vacuum state is generated by inserting the identity operator.

Suppose that we want to compute correlators $\left\langle \Psi|T \mathcal{O}_1(x_1)...\mathcal{O}_n(x_n)|\Psi \right\rangle$ of gauge-invariant operators $\mathcal{O}_i$ in a given initial state $\Psi$ ($T$ is time-ordering symbol). We can write a generating functional in the form
\begin{equation}
Z_{QFT}[J^I; C] = \int \!{\mathcal{D}\phi \exp\left( - i \int_C \!{dt}\int \!{d^{d-1} x \sqrt{-g} \left( \mathcal{L}_{QFT}[\phi] + J^I \mathcal{O}_I(\phi) \right)} \right)},
%\label{eq:}
\end{equation}
where $J^I$ are the sources coupling to gauge-invariant operators $\mathcal{O}_I$. The path integral is performed for the fields living on the contour $C$ in the complex time plane.

\begin{figure}
\centering
\includegraphics[width=0.8\textwidth]{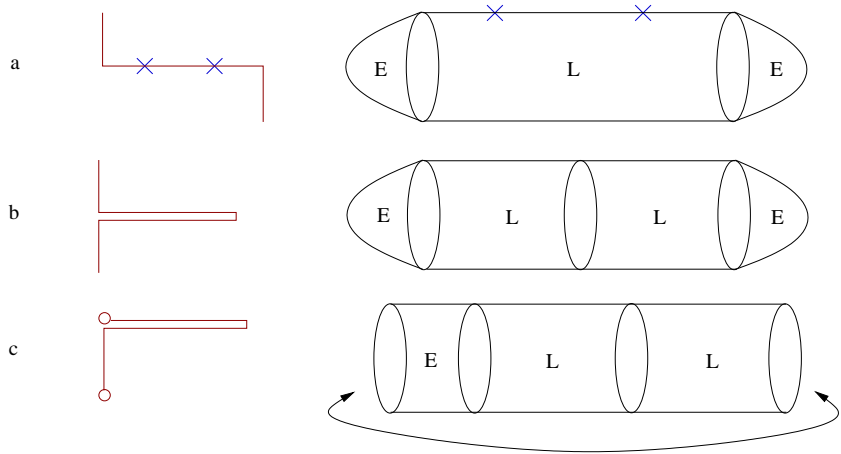}
\caption{Several possible contours can be used in the time plane to produce a) time-ordered correlator, b) Wightman function, c) thermal correlators. The figure is taken from \cite{Skenderis:2008dg}.}
\label{con}
\end{figure}

In fact, the choice of the contour $C$ determines, what kind of correlator we are calculating. Some examples of the contours are presented on the figure \ref{con}. For instance, for real-time thermal correlators one can use the closed Keldysh - Schwinger contour in figure \ref{con}c. The vertical segment now represents the thermal density matrix $\hat{\rho} = \exp(-\beta \hat{H})$, with $\beta = 1/T$. The points indicated by circle should be identified, and the thermal correlators should satisfy periodic / antiperiodic periodicity conditions for bosons / fermions. Depending on which of two vertical segments we put the sources we get the Keldysh-Schwinger matrix of thermal propagators.

\subsection{Prescription}
After short recapitulation of basic field theoretic facts we turn to the formulating of holographic prescription. The contour dependence discussed in the previous subsection should be reflected in the bulk. Within the saddle-point approximation we associate to a QFT contour $C$ a supergravity solution ('fill in' the QFT contour). The horizontal segments must be filled with Lorentzian solutions, while vertical segments - with Euclidean solutions. These segments are then glued together along bulk hypersurfaces that end on the corners of the contour. The entire manifold $M_C$ obtained in this way has a metric whose signature jumps at the corners.

Given this manifold $M_C$, the next step is to compute the corresponding supergravity on-shell action. This action is then identified with the generating functional of correlators in non-trivial states discussed in the previous subsection
\begin{equation}
Z_{QFT}[J^I; C] = \exp \left( i \int_{M_C} \!{d^{d+1} x \sqrt{-G} \mathcal{L}_{SUGRA}[\varphi, \varphi(u \rightarrow 0)=J^I]} \right),
\label{eq:pres}
\end{equation}
where $G$ is the determinant of the bulk metric and $J^I$ is the boundary value of the bulk field $\varphi$ dual to gauge-invariant operator $\mathcal{O}$. On Euclidean segments time is imaginary and after Wick rotation $t \rightarrow -i t$ one gets standart sign in front of the action. The sources $J^I$ that are localized on the conformal boundary of Euclidean segments are related to the initial and final state (identity operator corresponds to the vacuum). Whereas on the boundary of Lorentzian segments they correspond to the real physical sources and the $n$-point correlation functions can be produced via the functional differentiation with respect to them. In the bulk of this thesis we will be interested only in vacuum correlators, although this formalism can be applied also to correlators in non-trivial states.

\subsection{Matching Conditions and Corners}

Piecewise straight contours have corners, where either vertical segment meets horizontal one or two horizontal segments running in opposite directions join. These corners extend to hypersurfaces in the bulk. We impose following condition on them: the induced metric and all the fields and their conjugated momenta must be continuous across the corner. Note, that momenta are defined with respect to complexified time variable. These conditions give rise to the matching equations which allow us to find the unique correlation function corresponding to the given contour.

This matching condition can be justified in the following way. Imagine, that we have string theory on some manifold $M$. The generating functional is given by the path integral over all possible field configurations. This path integral can be written in a different way: split initial manifold $M$ in two pieces $M_1$ and $M_2$ along some hypersurface $S$. Then initial path integral can be replaced by the product of two path integrals over all possible field configurations on $M_1$ and $M_2$ with the given boundary value of the field times the integral over all possible boundary data, i.e. field configurations on the hypersurface $S$. The continuity of the fields is imposed by the fact that all fields must have the same value on the boundaries of $M_1$ and $M_2$. In saddle-point approximation the path integrals reduce to the exponentiated on-shell actions. Then perform the second saddle-point approximation with respect to the boundary data, i.e. vary on-shell actions with respect to the boundary value of the fields. Variation of the on-shell action with respect to the field gives precisely the conjugate momentum, i.e. momenta must also be continuous across the gluing surface $S$.

\subsection{Renormalization}
The fundamental relation (\ref{eq:pres}) is a bare relation, since both sides are generically infinite. On the QFT side there are UV divergences, but on the gravitational sides the divergences appear because of the infinite volume effects. To make this relation well-defined both sides need to be renormalized appropriately. This renormalization procedure is a priori more complicated then in Euclidean case.

In the Euclidean case the renormalization is done by introducing a set of local covariant counterterms. They are needed in order to make the on-shell action finite and the variational principle to be well-posed. In the Lorentzian setup there might appear new divergences. First, there is an additional non-compact direction: time. This difficulty is overcome by gluing Euclidean manifold near timelike infinities. Effectively it replaces dangerous part of the Lorentzian manifold by the radial boundary of Euclidean AdS, whose asymptotic structure is well known. The second and the last problem are the possible infinities at the corners. In principle, there can be new corner infinities which would require new counterterms. The absence of such is guaranteed by the matching conditions (compare \cite{Skenderis:2008dg}).

\subsection{Example}

\begin{figure}
\centering
\includegraphics{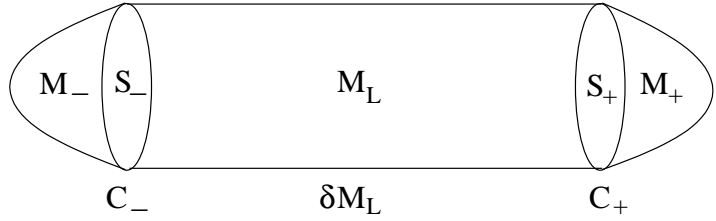}
\caption{The $CFT_2$ 'filled in' contour for the calculation of the time-ordered propagator. The figure is taken from \cite{Skenderis:2008dh}.}
\label{cigar}
\end{figure}

We are going to illustrate this formalism by an relatively easy low-dimensional AdS$_3$/CFT$_2$ example of scalar field (for more details consult \cite{Skenderis:2008dh}. Boundary CFT lives on the cylinder $S^1 \times \mathbf{R}$ (where $\mathbf{R}$ represents time direction) and hence we expect the spectrum to be discrete. We are going to compute time-ordered vacuum-to-vacuum correlator. We start with the contour in the time plane in figure \ref{Feynman}. The corners of the contour are two circles which we denote as $C_\pm$. The corresponding composed manifold consists of three pieces: a segment $M_L$ of
Lorentzian AdS$_{3}$ and two `caps' $M_\pm$ consisting of half of Euclidean AdS$_{3}$ (see figure \ref{cigar}). One can view these caps as providing a Hartle-Hawking wave function on the hypersurfaces $S_\pm$ (where $\partial S_{\pm} = C_\pm$).

By the AdS/CFT conjecture
\begin{multline}
\label{eq:ansatzrealtimeadscft}
\langle 0| T \exp \Big( - i \int_{\delta M_L} \! d^d x \sqrt{-g} \phi_{(0)} \op  \Big) |0 \rangle \\ = \exp\Big(i S_L[\phi_{(0)},\phi_-,\phi_+]
- S_E[0,\phi_-] - S_E[0,\phi_+]\Big).
\end{multline}
with $\delta M_L$ the conformal boundary of $M_L$ as in figure \ref{cigar}, $S_L[\phi_{(0)},\phi_-,\phi_+]$ the on-shell Lorentzian
action for $M_L$ that depends not only on $\phi_{(0)}$ but also on initial and final
data $\phi_{\pm}$, and $S_E[\phi_{(0,\pm)}, \phi_\pm]$ the
Euclidean on-shell actions on the half Euclidean spaces $M_{\pm}$
with sources $\phi_{(0,\pm)}$ and boundary condition $\phi_{\pm}$ at $S_\pm$.
In (\ref{eq:ansatzrealtimeadscft}) we set the
sources  $\phi_{(0,\pm)}$ to zero since we are interested in
vacuum-to-vacuum
correlators. Nonzero values for $\phi_{(0,\pm)}$ would correspond to changing the initial and/or final state, as it does in the CFT.

$\phi_\pm$ are fixed by imposing continuity of fields and conjugated momenta at the corners. Second one is equivalent to the stationarity of the on-shell action with respect to boundary values $\phi_\pm$:
\begin{equation}
\label{eq:secondmatching}
\frac{\delta}{\delta \phi_\pm} \Big( i S_L[\phi_{(0)},\phi_-,\phi_+]
- S_E[0,\phi_-] - S_E[0,\phi_+] \Big) = 0
\end{equation}
which should be read as an equation for $\phi_\pm$.

We now specialize to a free massive scalar $\Phi$. The relevant part of the supergravity action is:
\begin{equation}
\label{eq:actionscalar}
S = \frac{1}{2}\int \! d^3 x \sqrt{|G|} (- \partial_\mu \Phi \partial^\mu \Phi - m^2 \Phi^2).
\end{equation}
Dual operator $\mathcal O$ has conformal dimension $\Delta = 1 + \sqrt{1+m^2} = 1
+ l$ with $l \in \{0,1,2, \ldots\}$.

We take the metric for AdS$_3$ space in the form
\begin{equation}
ds^2 = - (r^2 + 1) dt^2 + \frac{dr^2}{r^2 + 1} + r^2 d\phi^2.
\end{equation}
The mode solutions to the equation equation of motion on this background are
$e^{-i\omega t + ik\phi}f(\omega,\pm k,r)$ with
\begin{align}
f(\omega,k,r) &=
C_{\omega k l} (1+r^2)^{\omega/2} r^{k}
F(\hat{\omega}_{kl},\hat{\omega}_{kl}-l;k+1;-r^2) \nonumber \\
&= r^{l-1}+ \ldots + r^{-l-1}\alpha(\omega,k,l) [\ln(r^2) + \beta(\omega,k,l)] + \ldots
\end{align}
where $\hat{\omega}_{kl} = (\omega + k + 1 + l)/2$, $C_{\omega k l}$
is a normalization factor chosen
such that the coefficient of the leading term equals 1
and in the last line we omitted terms of lower powers of $r$
and some terms polynomial in $\omega$ and $k$ (which would lead to contact terms in the 2-point function). Furthermore,
\begin{align}
%\begin{multline*}
\alpha(\omega,k,l)&=
(\hat{\omega}_{kl} -l)_l (\hat{\omega}_{kl} -k-l)_l/(l! (l-1)!)\, ,
\nonumber \\
%\frac{1}{\Gamma(l+1)\Gamma(l)}
%\frac{\Gamma(\frac{1}{2}(\omega + k + 1 +l))
%\Gamma(\frac{1}{2}(\omega - k + 1 +l))}{\Gamma(\frac{1}{2}(\omega + k + 1 -l))
%\Gamma(\frac{1}{2}(\omega - k + 1 -l))},
\beta(\omega,k,l) &= - \psi(\hat{\omega}_{kl}) -
\psi(\hat{\omega}_{kl} -\omega -l)\, ,
%\end{multline*}
\end{align}
where $(a)_n = \Gamma(a+n)/\Gamma(a)$ is the Pochhammer symbol
and $\psi(x)=d \ln \Gamma(x)/dx$ is the digamma function.
Note also that $f(\omega,k,r) = f(-\omega,k,r)$. Only the $f(\omega,k,r)$ with $k \geq 0$ are regular for $r\to 0$, so the modes we use below are of the form $e^{-i\omega t + ik\phi} f(\omega,|k|,r)$.

\begin{figure}
\centering
\includegraphics{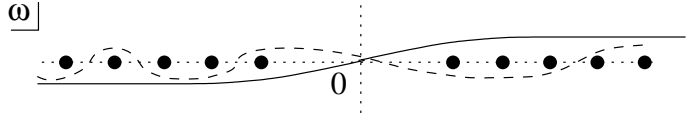}
\caption{There are many ways to define the integration contour in the $\omega$ plane. The figure is taken from \cite{Skenderis:2008dh}.}
\label{freq}
\end{figure}

We would now like to obtain the most general solution whose leading
asymptotic ($\sim r^{l-1}$ as $r \to \infty$) contain an arbitrary source
$\phi_{(0)}(t,\phi)$ for the dual operator.  Clearly, it will consist of non-normalizable mode with given asymptotic behavior plus eventually some normalizable modes. Thus our ansatz for the solution is
\begin{align}
\Phi(t,\phi,r) &= \frac{1}{4\pi^2}\sum_{k \in \mathbb Z} \int_C \! d\omega
\int \! d\hat t \int \! d \hat \phi e^{-i\omega (t - \hat t)
+ ik(\phi - \hat \phi)} \phi_{(0)}(\hat t, \hat \phi) f(\omega,|k|,r)
\nonumber \\
&\hspace{2cm}+ \sum_{\pm} \sum_{k \in \mathbb Z}
\sum_{n = 0}^{\infty} c_{nk}^\pm e^{-i\omega_{nk}^\pm t + ik\phi}g(\omega_{nk},|k|,r),
\label{eq:mostgeneralphil}
\end{align}
where $C$ represents
a contour in the complex $\omega$-plane which defines how do we go around the poles at:
\begin{equation}
\omega = \omega_{nk}^\pm \equiv \pm(2n + k + 1 + l) \, , \quad n \in \{0,1,2,\ldots\}.
\end{equation}
We are now completely free to specify any contour that circumvents
the poles (figure \ref{freq}). The difference between two different contours is a
sum over the residues:
\begin{align}
g(\omega_{nk},k,r) &= \oint_{\omega_{nk}} d\omega f(\omega_{nk},k,r) \nonumber \\
&\sim r^{-l-1} \alpha(\omega_{nk},k,l) \Big( \oint_{\omega_{nk}} d\omega \beta(\omega,k,l) \Big).
\end{align}
The $g(\omega_{nk},k,r)$ are the `normalizable modes'. Since a change of
contour can be undone by also changing the $c_{nk}^\pm$, let us fix
the contour to be the Feynman contour (solid line in figure \ref{freq}).

Now consider the solution on the `initial cap', so on the space
specified by the metric,
\begin{equation}
ds^2 = (r^2 + 1) d\tau^2  + \frac{dr^2}{r^2 +1} + r^2 d\phi^2
\end{equation}
with $-\infty < \tau \leq 0$, so that we have half of
Euclidean AdS space. On this segment there are no sources and only normalizable modes are allowed. Since the solution should vanish at $\tau
\to - \infty$,
the most general Euclidean solution contains only negative frequencies,
\begin{equation}
\Phi(\tau,\phi,r) = \sum_{n,k} d_{nk}^-
e^{-\omega_{nk}^- \tau + ik\phi}g(\omega_{nk},|k|,r)\, ,
\end{equation}
with thus far arbitrary coefficients $d_{nk}^-$.

We can now consider the matching at
$\tau = t = 0$, which will fix the initial data. From the continuity
$\Phi_L(0,\phi,r) = \Phi_E(0,\phi,r)$ we find, using orthogonality
and completeness of the $g(\omega_{nk},|k|,r)$ (for some more details see \cite{Skenderis:2008dh}):
\begin{equation}
\phi_{(0)}(\omega_{nk}^-,k) + c_{nk}^- + c_{nk}^+ = d_{nk}^-
\end{equation}
Equation (\ref{eq:secondmatching}) yields a relation between conjugate momenta,
\begin{equation}
- i \partial_t \Phi_L  = \partial_\tau \Phi_E\, .
\end{equation}
Substituting the solutions we find
\begin{equation}
- \omega_{nk}^- \phi_{(0)}(\omega_{nk}^-,k) -
\omega_{nk}^- c_{nk}^- -\omega_{nk}^+ c_{nk}^+ = - \omega_{nk}^- d_{nk}^-\, ,
\end{equation}
so that $c_{nk}^+ = 0$. Similarly, the matching to the out state
determines $c_{nk}^- = 0$, and indeed all the freedom in the
bulk solution is fixed. Had we chosen any other
contour in (\ref{eq:mostgeneralphil}), we would have found
nonzero values of some of the $c_{nk}^\pm$, effectively throwing us
back to the Feynman contour in figure \ref{freq}.

Finally, the two-point function is obtained from the
$r^{-l-1}$ term in the asymptotic expansion of
(\ref{eq:mostgeneralphil}) (with  $c_{nk}^\pm = 0$):
\begin{multline}
\langle 0| T \op(t,\phi) \op(0,0)  | 0 \rangle=\\
\frac{l}{4\pi^2 i} \sum_k \int_C \! d\omega e^{-i\omega t + ik\phi} \alpha(\omega,|k|,l)
\beta(\omega,|k|,l).
\end{multline}
with the contour $C$ being the same as for the bulk solution, thus
the standard Feynman prescription leading to time ordered correlators.
We emphasize again that $C$ was completely fixed by the matching to the
caps. Integrating over $C$ is equivalent to integrating over the real axis 
and shifting $\omega \to \omega(1+ i \epsilon)$. The Fourier transform of this 
expression then gives
%The Fourier transform is obtained by picking up residues and we obtain,
\begin{equation}
\langle 0| T \op(t,\phi) \op(0,0) |0 \rangle
= \frac{l^2/(2^{l+1} \pi)}{[\cos(t -i\epsilon t) - \cos(\phi)]^{l+1}}.
\end{equation}
This is the expected form for a time-ordered
two-point function on a cylinder and the normalization coefficient 
can be shown to agree with the standard AdS/CFT normalization of
2-point functions.

\chapter{Holographic Renormalization for Fermions on AAdS}

We begin with some general remarks concerning fermions on asymptotically AdS spacetimes \cite{Faulkner:2009wj, Iqbal:2009fd}.

A Dirac field $\psi$ in the bulk with charge $q$  is dual to a fermionic operator $\op$ in CFT of the same charge. $\op$ is a Dirac spinor for $d$ odd, and  a chiral spinor for $d$ even. In both cases the dimension of the boundary spinor $\op$ is half of that of $\psi$. For AdS space the conformal dimension $\Delta$ of $\op$  is given in terms of the mass $m$ of $\psi$ by $\Delta = \frac{d}{2} \pm m$. $\Delta$ cannot be negative, therefore for $m > d/2$ one has only one possibility: $\Delta = \frac{d}{2} + m$. For $0 < m < d/2$ there are two ways to quantize $\psi$ by imposing different boundary conditions at the boundary, which corresponds to two different
CFTs. Usual interpretation is that one of them is stable and another is unstable, i.e. there exists a deformation by some operator which makes it flow towards the stable theory \cite{Witten:2001ua}.

In the bulk of the thesis we will assume that the mass $m$ of the fermion is positive. Negative mass corresponds to the opposite chirality.

\section{Equation of Motion for Fermions on Lifshitz (AdS) Spacetime}

AdS spacetime is well known to be particular case of the so called Lifshitz spacetime which in turn is defined by the metric

\begin{equation}
ds^2 = - \frac{dt^2}{u^{2 z}} + \frac{du^2 + dx^2}{u^2}.
\label{metric}
\end{equation}
$z$ is called the dynamical exponent. For the case of $z=1$ the metric (\ref{metric}) reduces to the AdS space. We derive the equation of motion on general Lifshitz spacetimes. Later we will specialize to particular cases.
We consider the quadratic part of the action
\begin{equation}
S = \int \! d^{d+1}x \sqrt{-g} i (\overline{\Psi} \Gamma^{M} D_M \Psi - m \overline{\Psi} \Psi)+ S_{\text{bdy}},
\label{eq:action}
\end{equation}
where
\begin{equation}
D_{M} = \partial_{M} + \frac{1}{4} (\omega_{M})_{AB}[\gamma^{A}, \gamma^{B}]
%\label{eq:}
\end{equation}
is the covariant derivative and $(\omega_{M})_{AB} $ are the spin connection coefficients. Letters from the beginning of the alphabet denote the tangent frame indices and those from the middle of the alphabet - spacetime indices. The \(\gamma^{A}\) satisfy the Clifford algebra \( \left\{\gamma^{A}, \gamma^{B}\right\} = 2 \eta^{AB}\). $S_{\text{bdy}}$ is the boundary action needed for the variational principle to be well posed. It is not important for the moment, since it does not contribute to the bulk dynamics. We will construct it in the next section. Non-vanishing spin connection coefficients are \( (\omega_t)_{tu} = z/u^z\) and $(\omega_x)_{xu} = -1/u $. Different gamma-matrices are related through the inverse vielbein $\Gamma^{M} = e^{M}_{A} \gamma^{A}$. %For convenience let us introduce the projection operator $\Pi_{\pm} = \frac{1}{2}(1 \pm \gamma^{u})$. Then $\Psi_{\pm}$ satisfy $\Pi_{\pm}\Psi_{\pm} = \pm \Psi_{\pm} $.
We get the Dirac operator
\begin{align}
\Gamma^{M}D_{M} &= u^z \gamma^{t} (\partial_{t} + \frac{1}{4}\frac{z}{u^z}[\gamma^{t},\gamma^{u}]) + u\gamma^{u}\partial_{u} + u \gamma^{i}(\partial_i -  \frac{1}{4}\frac{1}{u}[\gamma^{i},\gamma^{u}]) \nonumber \\
{} &= u^z \gamma^{t} \partial_{t} + u \gamma^{i} \partial_{i} + u \gamma^{u} \partial_{u} - \frac{d+z-1}{2}\gamma^{u}
%\label{eq:}
\end{align}
After Fourier transforming the spinor in boundary directions $ \Psi = e^{i \omega t - i k x} \psi(u)$ the Dirac equation becomes
\begin{equation}
(i \omega u^z \gamma^t + i k u \gamma^i + u \gamma^u \partial_u - \frac{d+z-1}{2} \gamma^u - m)\psi =0.
\label{eq:Dirac}
\end{equation}

Applying $(\Gamma^{M}D_{M})^2 - m^2$ to $\Psi(u)$ yields
\begin{align}
&\Big( u^2 \partial_u^2 - (d+z-1)u \partial_u + \big[ (z-1)i \omega u^z \gamma^u \gamma^t + \omega^2 u^{2 z} - \vec{k}^2 u^2 \nonumber \\ &\hspace{2cm}+ \left( \frac{d+z}{2}\right)^2 - \frac{1}{4} -m^2 +m \gamma^u \big] \Big) \psi(u) =0.
\label{eom}
\end{align}
We were not able to find the derivation of this result in the literature. One gets the equation of motion for Euclidean signature just by replacing $\omega^2 \rightarrow - \omega^2$.
Now there are two interesting cases for which we can solve this equation analytically. First, on pure AdS ($z=1$) the term containing the product $\gamma^u \gamma^t$ vanishes identically and we can define the Weyl projector $\Pi_{\pm} = \frac{1}{2}(1 \pm \gamma^{u})$. Then $\Psi_{\pm} = \Pi_{\pm} \Psi $ satisfy $\gamma^u \Psi_{\pm} = \pm \Psi_{\pm} $. The current chapter will be devoted mostly to this case. Another solvable case is $z=2$ and $m = 0$. Then we can decompose the Dirac spinor using $\Pi_{\pm} = \frac{1}{2}(1 \pm \gamma^{u} \gamma^t)$. The solution and analysis of this case is given in chapter 4.

\section{Boundary Term} \label{boundary_term}
The AdS/CFT in its weakest form is based on the stationary phase approximation for supergravity, i.e. we evaluate the action $S_{SUGRA}$ on-shell. For the stationary phase approximation it is crucial that the classical solution is indeed the stationary point for the action. For the spacetimes with boundaries this observation leads to some important consequences. Due to the possible boundary terms classical solution does not necessary is a stationary point. The point is that if even the variation of the bulk action on the solution vanishes, the variation of the boundary action can be different from zero. This problem can be cured by adding appropriate boundary term to the bulk action. For the spin - $2$ field this term is called Hawking - Gibbons term \cite{Gibbons:1976ue}. Now we are going to show how one constructs appropriate boundary action for the fermionic field \cite{Henneaux:1998ch, Iqbal:2009fd}.

We begin with the Dirac action (\ref{eq:action}) which we write here once again (in Euclidean signature) for convenience
\begin{equation}
S_{\text{bulk}} = -\int \!{d^{d+1} x \sqrt{g} \left( \overline{\Psi} D \!\!\!\! /  \Psi  - m \overline{\Psi} \Psi \right)}.
%\label{eq:}
\end{equation}
Variating this action and using the equations of motion for $\Psi$ and $\overline{\Psi}$ we get
\begin{align}
\delta S_{\text{bulk}} &= -\int \!{d^{d+1} x \sqrt{g} \left(  \delta \overline{\Psi} D \!\!\!\! / \Psi +\overline{\Psi} D \!\!\!\! / \delta \Psi - m \delta \overline{\Psi} \Psi - m \overline{\Psi} \delta \Psi \right)} \nonumber \\ 
&= -\int_0^{\infty} \!{du}\int \!{d^{d} x \sqrt{g} \left( \overline{\Psi} D \!\!\!\! / \delta \Psi - m \overline{\Psi} \delta \Psi \right)} \nonumber \\
&= \int \!d^{d} x \sqrt{g_{\text{induced}}} \overline{\Psi} \gamma^u \delta \Psi \nonumber \\
&= \int \!{d^d x \sqrt{g_{\text{induced}}} \left( \overline{\Psi}_- \delta \Psi_+ - \overline{\Psi}_+ \delta \Psi_- \right)},
\end{align}
where we have used projectors $\Pi_{\pm} = \frac{1}{2}(1 \pm \gamma^{u})$ to define $\Psi_{\pm} = \Pi_{\pm} \Psi$.

As we will see in the next section ${\Psi}_-$ and ${\Psi}_+$ are not independent. In fact ${\Psi}_+$ can be expressed in terms of ${\Psi}_-$, i.e. we are not allowed to vary ${\Psi}_+$ freely. In other words we must set $\delta {\Psi}_+ = 0$.

Now it is easy to see that $\delta S_{\text{bulk}}$ is itself the variation of a surface term at the boundary
\begin{equation}
\delta S_{\text{bulk}} = - \delta S_{\text{bdy}},
%\label{eq:}
\end{equation}
with
\begin{equation}
S_{\text{bdy}} = \int \!d^d x \sqrt{g_{\text{induced}}}  \overline{\Psi}_+ \Psi_-.
%\label{eq:}
\end{equation}

This boundary action $S_{\text{bdy}}$ must be added to the Dirac action in order to make variational principle well-posed. 

For the Lifshitz spacetimes the derivation is completely analogous and thus the boundary term has the same form.

\section{Euclidean Signature}
We begin by reviewing the renormalization procedure for fermions on the AdS with Euclidean signature \cite{Iqbal:2009fd, Ammon:2010pg}. Here we shall already see many important features which were not relevant for the bosons. Firstly, the equation of motion for the fermions is the first order equation. Because of it we should pose the Dirichlet problem particularly carefully. This complication is related to the other obvious problem: fermions in the bulk and those in the boundary theory have different numbers of components. Secondly, to set the variational action principle for the fermions in the bulk we must add a boundary term (see section \ref{boundary_term}), which will guarantee, that the action is extremized on the equation of motion \cite{Henneaux:1998ch}.

The classical AdS/CFT prescription says that the on-shell bulk action is the generator of connected correlators in the boundary theory:
\begin{equation}
\left\langle \exp[\int \!{d^d x \overline{\chi} O + \overline{O} \chi}]\right\rangle = e^{-S_{SUGRA}[\chi,\overline{\chi}]},
%\label{eq:}
\end{equation}
where $\chi$ and $\overline{\chi}$ are the boundary values of the bulk fermions and $S_{SUGRA}$ must be evaluated on the solution to the equation of motion (saddle point approximation). Note, that so far it is very formal equation, since generically both sides are infinite. In order to be able to extract finite result we must perform renormalization. But so far we shall work formally, as if everything is finite and well-defined.

Taking on both sides the functional derivative with respect to $\chi$ we get
\begin{equation}
\left\langle \overline{O} \right\rangle = -\frac{\delta S}{\delta \chi} = -\Pi_{\chi},
%\label{eq:}
\end{equation}
where $\Pi_{\chi}$ is the momentum conjugate to $\chi$ \cite{Papadimitriou:2004ap, Papadimitriou:2004rz}. The $2$-point function is given by
\begin{equation}
\left\langle  {O} \overline{O}\right\rangle = \frac{\delta^2 S}{\delta \chi \delta \overline{\chi}}.
%\label{eq:}
\end{equation}
\begin{comment}
How do we define the fermionic Euclidean propagator (or fermionic Euclidean 2-point function)? The answer is given by the linear response theory:
\begin{equation}
\left\langle \overline{O} \right\rangle = G(k) \gamma^t \chi.
%\label{eq:}
\end{equation}
\end{comment}
Equivalently, the Euclidean propagator is given by the matrix relating the (renormalized) 1-point function of the dual boundary fermionic operator $O$ and the source $\chi$
\begin{equation}
\left\langle \overline{O} \right\rangle = G(k) \gamma^t \chi.
%\label{eq:}
\end{equation}
These in turn can be identified with the leading coefficients in the power expansions of the normalizable and non-normalizable mode correspondingly. $\gamma^t$ appears because $ G = \left\langle O O^{\dagger} \right\rangle = \left\langle O \overline{O}\right\rangle \gamma^t $.

After giving the rough idea let's look how it really works on the example: fermions on asymptotically AdS. The equation of motion we get by plugging $z=1$ in the (\ref{eom}) and replacing $\omega^2$ by $-\omega^2$ (because of the change in signature)

\begin{equation}
\left[ \partial^{2}_{u} - \frac{d}{u} \partial_{u} + \frac{1}{u^2} \left( -m^2 \pm m + \frac{d^2}{4} + \frac{d}{2}\right) - k^2\right]\psi_{\pm} = 0,
%\label{eq:}
\end{equation}
where we have introduced $k^2 = \omega^2 + \vec{k}^2$ and defined $\psi_{\pm} = \frac{1}{2} (1 \pm \gamma^u) \psi$. The general solution for $m$ not a half-integer is 
\begin{equation}
\psi_{\pm} = u^{\frac{d+1}{2}} \lbrace C^{\pm}_1(k) I_{m\mp 1/2}(k u) + C^{\pm}_2(k) I_{-m\pm 1/2}(k u) \rbrace.
\label{eq:sol}
\end{equation}

When $m$ is a half-integer we need to introduce the modified Bessel function of the second kind $K$ as a second linearly independent solution and the general solution takes the form
\begin{equation}
\psi_{\pm} = u^{\frac{d+1}{2}} \lbrace C^{\pm}_1(k) I_{m\mp 1/2}(q u) + C^{\pm}_2(k) K_{m\mp 1/2}(q u) \rbrace.
\end{equation}

Using the series expansion of modified Bessel functions (Appendix A) we find the leading behavior of (\ref{eq:sol}) 
\begin{align}
\psi_{+} = c^+_1 (k) u^{\frac{d}{2} + m} + c^+_2 (k) u^{\frac{d}{2} - m + 1} \\
\psi_{-} = c^-_1 (k) u^{\frac{d}{2} + m + 1} + c^-_2 (k) u^{\frac{d}{2} - m}
%\label{eq:sol}
\end{align}
The questions arises: how should we impose boundary conditions. Naively, we could impose Dirichlet boundary conditions on both projections $\Psi_{\pm}$. But in this case we would fix the solution uniquely and generically it will not be regular in the bulk. The right thing to do is to impose first the regularity condition in the bulk, then solve for $\Psi_{\pm}$ and recognize the source as the leading coefficient of the non-normalizable mode.
We immediately see that near the boundary the dominant term has coefficient $c^-_2$. Thus, it corresponds to the source on the CFT side and we should impose the boundary condition $c^-_2 \sim \chi$. The normalizable mode of $\psi_+$ goes with the $c^+_1 (k)$ coefficient (being the response of the dual operator $O$). We want to find the matrix which relates $c^+_1 (k)$ and $c^-_2 (k)$.

\begin{comment}
Plugging these solutions back into the equation of motion \ref{eq:Dirac} we find that the coefficients are related by
\begin{align}
c^-_1 (k) = \frac{1}{2 m + 1} i \not{k} c^+_1 (k),\\
c^+_2 (k) = \frac{1}{2 m - 1} i \not{k} c^-_2 (k)
\label{eq:rel}
\end{align}
\end{comment}

Now we construct the on-shell action \cite{Ammon:2010pg}. Let us consider $m$ not half-integer. We have
\begin{align}
\Psi_{\pm} &= e^{- i \omega t + i \vec{k} \vec{x}} u^{\frac{d+1}{2}} \lbrace C^{\pm}_1(k) I_{m\mp 1/2}(k u) + C^{\pm}_2(k) I_{-(m\mp 1/2)}(k u) \rbrace \nonumber \\ &= c_1^{\pm} u^\frac{d}{2} + m \mp \frac{1}{2} + \frac{1}{2} (1 + s_a^{\pm}(u,k)) \nonumber \\ & \hspace{2cm}+ c_2^{\pm} u^\frac{d}{2} - m \pm \frac{1}{2} + \frac{1}{2} (1 + s_b^{\pm}(u,k)),
\end{align}
where we redefined $C$'s multiplied with some factors by $c$'s and we have defined the series
\begin{align}
s_a^{\pm}(u,k) &\equiv \sum_{j=1}^{\infty}{a_j^{\pm}(m) (-k^2)^j u^{2j}}, \nonumber \\
a_j^{\pm}(m) &\equiv \frac{(-1^j)}{j! 2^{2 j}} \frac{\Gamma(1+(m\mp \frac{1}{2}))}{\Gamma(j+1+(m\mp \frac{1}{2}))}.
%\label{eq:}
\end{align}
$s_b$ and $b_j$ are defined similarly, but with $(m\mp \frac{1}{2}) \rightarrow - (m\mp \frac{1}{2})$.
We write $\Psi_+$ and $\Psi_-$ separately
\begin{align}
\Psi_+ &= c_1^+ u^{\frac{d}{2} + m}(1 + s_a^{+}(u,k)) +c_2^{+} u^{\frac{d}{2} - m + 1} (1 + s_b^{+}(u,k)), \\
\Psi_- &= c_1^- u^{\frac{d}{2} + m +1}(1 + s_a^{-}(u,k)) +c_2^{-} u^{\frac{d}{2} - m} (1 + s_b^{-}(u,k)),
%\label{eq:}
\end{align}
and identify the source as the term multiplying the $c_2^-$ coefficient and the responce as the term multiplying $c_1^+$ coefficient (when $m > 1/2$).
The coefficients $c_{1,2}^{\pm}$ are not actually independent. If we plug the solution back into (\ref{eq:Dirac}) (with $z=1$) and collect powers of $u$ we get (if $m$ is not integer) 
\begin{align}
0 & = [(-2 m + 1) c_2^+ + i \not{k} c_2^-] u^{\frac{d}{2} -m +1} \\ & \hspace{3cm}+ [(-2 m + 1) c_1^- + i \not{k} c_1^+] u^{\frac{d}{2} +m +1} + ...
%\label{eq:}
\end{align}
where we have generalized $k \gamma^i \rightarrow \not{k}$. It follows that
\begin{align}
c_1^- &= \frac{1}{2 m + 1} i \not{k} c_1^+, \\
c_2^+ &= \frac{1}{2 m - 1} i \not{k} c_2^-
\label{eq:rel}
\end{align}
(now it is again clear, that we were not allowed to impose boundary conditions on both $\psi_+$ and $\psi_-$).

Now we repeat the same exercise for half-integer $m$.
\begin{align}
\Psi_{\pm} &= e^{- i \omega t + i \vec{k} \vec{x}} u^{\frac{d+1}{2}} \lbrace C^{\pm}_1(k) I_{m\mp 1/2}(k u) + C^{\pm}_2(k) K_{(m\mp 1/2)}(k u) \rbrace \nonumber \\ &= c_1^{\pm} u^{\frac{d}{2} + m \mp \frac{1}{2} + \frac{1}{2}} \ln u (1 + s_a^{\pm}(u,k)) + c_2^{\pm} u^{\frac{d}{2} - m \pm \frac{1}{2} + \frac{1}{2}} (1 + s_d^{\pm}(u,k)).
\label{log:sol}
\end{align}
Note that we are using the units in which the radius of $AdS$ is equal to 1. The argument of the logarithm includes factors of the radius to render them dimensionless. The $d_j$ coefficients are defined differently from $a_j$ and $b_j$, but the specific expressions for them is not important for us at the moment. Written separately
\begin{align}
\Psi_+ &= c_1^+ u^{\frac{d}{2} + m}\ln u (1 + s_a^{+}(u,k)) +c_2^{+} u^{\frac{d}{2} - m + 1} (1 + s_b^{+}(u,k)), \\
\Psi_- &= c_1^- u^{\frac{d}{2} + m +1} \ln u (1 + s_a^{-}(u,k)) +c_2^{-} u^{\frac{d}{2} - m} (1 + s_d^{-}(u,k)).
%\label{eq:}
\end{align}
Again, the coefficients are not independent. In fact, when $m \neq 1/2$ they are related in the same way as for not half-integer $m$. For $m = 1/2$ one gets 
\begin{equation}
c_1^+ = - i \not{k} c_2^-.
%\label{eq:}
\end{equation}

Now we turn to the evaluation of the on-shell action.
As already mentioned the bulk term vanishes when evaluated on a solution. The nonzero contribution comes from the boundary term $S_{\text{bdy}}$. We split $S_{\text{bdy}}$ into two terms
\begin{equation}
S_{\text{bdy}} = S_{\text{var}} + S_{CT},
%\label{eq:}
\end{equation}
where $S_{\text{var}}$ are terms required for the variational principle and $S_{CT}$ includes counterterms which will cancel the divergences. As we already know
\begin{equation}
S_{\text{var}} = \int \!{d^d x \sqrt{\gamma} \overline{\Psi}_+ \Psi_-}
%\label{eq:}
\end{equation}
where the integration is over $u=\epsilon$ surface and $\gamma$ is the determinant of the induced metric. 

For $m$ not half-integer
\begin{align}
S_{\text{var}} = \int \! d^d x \frac{1}{\epsilon^d} ( &\overline{c}_1^+ c_2^- \epsilon^d (1 + f_{a^+ b^-})\nonumber \\ &+\overline{c}_2^+ c_1^- \epsilon^{d+2} (1 + f_{b^+ a^-}) \nonumber \\&+\overline{c}_1^+ c_1^- \epsilon^{d+2m+1} (1 + f_{a^+ a^-}) \nonumber \\&+  \overline{c}_2^+ c_2^- \epsilon^{d - 2 m + 1} (1 + f_{b^+ b^-}) ),
%\label{eq:}
\end{align}
where we have defined
\begin{equation}
 f_{a^+ b^-} = s_a^+(\epsilon,k)+ s_b^-(\epsilon,k) + s_a^+(\epsilon,k) s_b^-(\epsilon,k)-,
\end{equation}
and similarly for $f_{a^+ a^-}$, $f_{b^+ a^-}$, $f_{b^+ b^-}$, all of which are the power series in $\epsilon^2$ starting with $\epsilon^2$. We now see that only the fourth term can diverge if $m > 1/2$. We want to rewrite $S_{\text{var}}$ in terms of the $c_1^+$ and $c_2^-$ (response and source). Using (\ref{eq:rel}) we get 
\begin{equation}
S_{\text{var}} = \int \!{d^d x \frac{1}{\epsilon^d} [ \overline{c}_1^+ c_2^- \epsilon^d  \frac{1}{2 m - 1}\overline{c}_2^- i \not{k} c_2^- \epsilon^{d - 2 m + 1}(1 + f_{b^+ b^-}) + O(\epsilon^{d+2})]}.
%\label{eq:}
\end{equation}
After having isolated the divergences we must write an $S_{CT}$ which must cancel the divergences and has to respect the symmetries of the theory, i.e. must be covariant in the source - boundary value of $\Psi_-$. The appropriate $S_{CT}$ is given by
\begin{align}
S_{CT} &= \int \!{d^d x \sqrt{\gamma} \sum_{j = 0}^{\infty}{\alpha_j(m) \overline{\Psi}_- \not{\partial}_{\epsilon} \Box^j_\epsilon \Psi_-}} \nonumber \\ &= \int \!{d^d x \frac{1}{\epsilon^d} \sum_{j = 0}^{\infty}{\epsilon^{1+2j} \alpha_j(m) \overline{\Psi}_- \not{\partial}\Box^j\Psi_-}},
%\label{eq:}
\end{align}
where $\not{\partial}_{\epsilon} = \epsilon \not{\partial}$ (the power of $\epsilon$ comes from the inverse vielbein evaluated at $u = \epsilon $) and $\Box^j_\epsilon$ is some power $j$ of the scalar Laplacian $\Box_\epsilon$ on the $u=\epsilon$ surface, which in our case is simply $\Box_\epsilon = \epsilon^2 \partial^2$. Coefficients $\alpha_j(m)$ are still to be determined. When we take $\Psi = e^{i k x} \psi$ and plug in the solution, the counterterms become
\begin{align}
S_{CT} &= \int \!{d^d x \frac{1}{\epsilon^d} \sum_{j = 0}^{\infty}{\epsilon^{1+2j} \alpha_j(m) \overline{\Psi}_- \not{k}(-k^2)^j\Psi_-}} \nonumber \\
&= \int \! d^d x \frac{1}{\epsilon^d} \sum_{j = 0}^{\infty}{\epsilon^{1+2j} \alpha_j(m)}
( \epsilon^{d + 2 m + 2} \overline{c}_1^- i \not{k} (-k^2)^j c_1^- (1 + f_{a^-a^-}) \nonumber \\
& \hspace{4.6cm}+ \epsilon^{d + 1} \overline{c}_1^- i \not{k} (-k^2)^j c_2^- (1 + f_{a^-b^-}) \nonumber \\
& \hspace{4.6cm}+ \epsilon^{d + 1} \overline{c}_2^- i \not{k} (-k^2)^j c_1^- (1 + f_{b^-a^-}) \nonumber \\
& \hspace{4.6cm}+ \epsilon^{d - 2 m} \overline{c}_2^- i \not{k} (-k^2)^j c_2^- (1 + f_{b^-b^-})).
%\label{eq:}
\end{align}
The coefficients $\alpha_j(m)$ are determined by the requirement that the last term must cancel potential divergences in $S_{\text{var}}$, i.e.
\begin{equation}
\frac{1}{2 m -1} (1 + f_{b^+b^-}) + \sum_{j = 0}^{\infty}{ \alpha_j(m) (-\epsilon^2 k^2)^j (1 + f_{b^-b^-})}. 
%\label{eq:}
\end{equation}
must vanish order by order in $-\epsilon^2 k^2$ up to order $\epsilon^{2m-1}$. From this equation all the $\alpha_j(m)$ can be determined recursively and thus $S_{CT}$ is fixed. Explicit expressions for the first of them one can find in \cite{Ammon:2010pg}.

The similar story for half-integer $m$ can be found in \cite{Ammon:2010pg}.

Now we calculate the 2-point function for the easiest case: pure AdS with not half-integer $m$. For that we impose the regularity condition on the (\ref{eq:sol}).
\begin{comment}
The normalizable mode of $\psi_+$ goes with the $c^+_1 (k)$ coefficient (being the response of the dual operator $O$). We want to find the matrix which relates $c^+_1 (k)$ and $c^-_2 (k)$. This is easily done by requiring regularity of the solution in the bulk and using (\ref{eq:rel}).
\end{comment}
Regularity is achieved only if $C^+_1 (k) = - C^+_2 (k)$. Note, that $I_n - I_{-n} \sim K_n$. This results in
\begin{align}
c^+_1 (k) &= \frac{1}{\Gamma(m + 1/2)} \left(\frac{k}{2}\right)^{m - 1/2} C^+_1 (k), \\
c^+_2 (k) &= \frac{1}{\Gamma(-m + 3/2)} \left(\frac{k}{2}\right)^{-m + 1/2} C^+_2 (k).
\end{align}
Collecting last $2$ equations together with (\ref{eq:rel}) we find the Euclidean propagator
\begin{equation}
G(k) = \frac{\Gamma(-m+1/2)}{\Gamma(m-1/2)} \left(\frac{k}{2}\right)^{2 m -1} \frac{1}{2 m -1} i \not{k}.\gamma^t
\end{equation}

For half-integer $m$ we pick $K_n$ immediately as the regular solution and get similar result.

\section{Lorentzian Signature}
Let's think about renormalization. If there are no sources on Euclidean segments of QFT (as is the case for the vacuum correlators) then only normalizable modes are allowed on Euclidean AdS. These give only finite contribution to the action (Dirac action vanishes obviously and $S_{\text{var}}$ is finite). Analysis on the conformal boundary of Lorentzian AdS is exactly the same as for the Euclidean one (we have only to distinguish between spacelike and timelike momenta). So, the only source of possible divergences are the hypersurfaces along which Euclidean and Lorentzian segments are glued together. The absence of these divergences is guaranteed by the matching conditions (compare \cite{Skenderis:2008dg}).

\subsection{Feynman Propagator}\label{}
The equation of motion on the Lorentzian segment we get by plugging $z=1$ in (\ref{eom}):
\begin{equation}
\left[ \partial^{2}_{u} - \frac{d}{u} \partial_{u} + \frac{1}{u^2} \left( -m^2 \pm m + \frac{d^2}{4} + \frac{d}{2}\right) - q^2\right]\psi_{\pm} = 0
%\label{eq:}
\end{equation}
This is the same equation as one gets for Euclidean signature of AdS. The only difference is that now $q^2 = - \omega^2 + \vec{k}^2$ and one has to distinguish between spacelike and timelike momenta.

Next, we want to discuss the solution to this equation of motion. The solution for spacelike momenta when $m$ is not a half-integer is
\begin{equation}
\psi_{\pm} = u^{\frac{d+1}{2}} \lbrace C^{\pm}_1(k) I_{m\mp 1/2}(q u) + C^{\pm}_2(k) I_{-(m\mp 1/2)}(q u) \rbrace,
\end{equation}
where $I$ is a modified Bessel functions of the first kind and $C^{\pm}_1$, $C^{\pm}_2$ are spinors of definite chirality. We behold both solutions, since we are interested not only in pure AdS background, but also in asymptotically AdS, i.e. both solutions can play a role depending on the condition in the interior of the bulk. 

When $m$ is a half-integer we need to introduce the modified Bessel function of the second kind $K$ as a second linearly independent solution and the general solution takes the form
\begin{equation}
\psi_{\pm} = u^{\frac{d+1}{2}} \lbrace C^{\pm}_1(k) I_{m\mp 1/2}(q u) + C^{\pm}_2(k) K_{m\mp 1/2}(q u) \rbrace.
\end{equation}
 To get the solution for timelike momenta we analytically continue the solution for spacelike momenta to the case of imaginary arguments and get (for $m$ - half-integer)
\begin{equation}
\psi_{\pm} = u^{\frac{d+1}{2}} \lbrace C^{\pm}_1(k) J_{m\mp 1/2}(q u) + C^{\pm}_2(k) Y_{m\mp 1/2}(q u) \rbrace,
\end{equation}
where $J$ and $Y$ are Bessel functions. 
From the series expansions of Bessel functions (Appendix A) we see, that $J_n$ ($I_n$) corresponds to the normalizable mode while $Y_n$ ($K_n$) corresponds to the source. Deep in bulk these functions behave as
\begin{equation}
u^{\frac{d+1}{2}}J_{m \mp 1/2}(qu) \approx \sqrt{\frac{2}{\pi q}} u^{d/2} \cos \left(qu - \frac{(m \mp 1/2) \pi}{2} -\frac{\pi}{4}\right),
\end{equation}
\begin{equation}
u^{\frac{d+1}{2}}Y_{m \mp 1/2}(qu) \approx \sqrt{\frac{2}{\pi q}} u^{d/2} \sin \left( qu - \frac{(m \mp 1/2) \pi}{2} -\frac{\pi}{4} \right)
\end{equation}
\begin{equation}
u^{\frac{d+1}{2}}K_{m \mp 1/2}(qu) \approx \sqrt{\frac{\pi}{2 q}} u^{d/2} e^{-q u}
\end{equation}
which shows, that for timelike momenta no linear combination of the solutions remains finite as $u\rightarrow\infty$, i.e. any solution that does remain finite as $u\rightarrow\infty$ should be obtained as an infinite sum over the modes. For space-like momenta regularity in the bulk selects $K_n$ as the only possible solution. But note, that this solution is not normalizable.

After we understood the structure of the solution on the Lorentzian bulk $M_1$, let us return to the prescription of Balt C. van Rees and Skenderis. Consider the Euclidean manifolds $M_0$ and $M_2$ with time coordinates $-\infty < \tau < 0$ and $0 < \tau < \infty$, respectively (compare figure \ref{Feynman}). The mode solutions on $M_0$ and $M_2$ are obtained by the usual replacement $t \rightarrow - i \tau$ in the Lorentzian modes. Physically, we do not have any sources on these segments, thus the solutions on $M_0$ and $M_2$ should be purely normalizable. Furthermore, only negative (positive) frequencies are allowed on $M_0$ ($M_2$). Explicitely the solutions on these segments are the linear combinations of
\begin{align}
e^{|\omega| \tau_0 + i \vec{k} \vec{x}} u^{\frac{d+1}{2}} J_{m \mp 1/2}(q u)  \textrm{\qquad on $M_0$} \nonumber \\
e^{-|\omega| \tau_2 + i \vec{k} \vec{x}} u^{\frac{d+1}{2}} J_{m \mp 1/2}(q u) \textrm{\qquad on $M_2$}
\end{align}

Next, we construct the mode which extends over all the segments. On the Lorentzian segment (by analogy with the Euclidean case) we try
\begin{equation}
\Psi^{\pm}_1(t, \vec{x}, u) = \frac{1}{(2 \pi)^d} \int _{C}\! d\omega \int \! d\vec{k} e^{- i \omega t + i \vec{k} \vec{x}}\frac{q_{\epsilon}^{m \mp 1/2}}{2^{m-1 \mp 1/2} \Gamma(m \mp \frac{1}{2})} u^{\frac{d+1}{2}} K_{m \mp 1/2}(q_{\epsilon}u)
\label{eq:propagator}
\end{equation}
Note, that we still have to specify the integration contour $C$, since the Bessel functions $Y_n$ and $K_n$ have branch cuts for integer index $n$. To understand, what is happening, note that $K_{m \mp 1/2}(qu)$ is unambiguously defined for spacelike momenta $q^2 = - \omega^2 + \vec{k}^2 >0$. For timelike momenta $q^2 < 0$ we have to consider branch cuts. We define the square root $q_\epsilon = \sqrt{-\omega^2 + \vec{k}^2 - i \epsilon}$ (just above the negative real axis). This choice (as we shall see later) corresponds to the Feynman $i\epsilon$ prescription for the propagator.
To check that (\ref{eq:propagator}) is finite in the bulk ($u \rightarrow \infty$) we perform the integration by deforming the contour and integrating along the branch cut. The result is \cite{Skenderis:2008dg}
\begin{equation}
\Psi^{\pm}_1(t, \vec{x}, u) = i \pi^{-d/2} \frac{\Gamma(m \mp 1/2) \Gamma(d + m \mp 1/2)}{2^{2m \mp 1}} \frac{u^{d + m + 1/2 \mp 1/2}}{(- t^2 + \vec{x}^2 + u^2 + i \epsilon)^{d + m \mp 1/2}},
\label{pos_prop}
\end{equation}
which is obviously finite for large $u$ (but note, that asymptotic behavior differs from that of scalar field). By analogy with \cite{Skenderis:2008dg} we easily find the extensions to the Euclidean segments
\begin{align}
&\Psi^{\pm}_0(\tau_0, \vec{x}, u) = i \pi^{-d/2} \frac{\Gamma(m \mp 1/2) \Gamma(d + m \mp 1/2)}{2^{2m \mp 1}} \nonumber \\ &\hspace{3.5cm} \times \frac{u^{d + m + 1/2 \mp 1/2}}{(- (-T + i \tau_0)^2 + \vec{x}^2 + u^2 + i \epsilon)^{d + m \mp 1/2}}, \\
&\Psi^{\pm}_2(\tau_2, \vec{x}, u) = i \pi^{-d/2} \frac{\Gamma(m \mp 1/2) \Gamma(d + m \mp 1/2)}{2^{2m \mp 1}} \nonumber \\ &\hspace{3.5cm} \times \frac{u^{d + m + 1/2 \mp 1/2}}{(- (T - i \tau_2)^2 + \vec{x}^2 + u^2 + i \epsilon)^{d + m \mp 1/2}},
\end{align}
satisfying matching conditions. We will show how to find these modes in momentum space in section \ref{thermal}. $i \epsilon$ insertions are needed on the on the initial and final hypersurfaces given by $\tau_0 = 0$ and $\tau_2 =0$. Obviously the matching conditions are satisfied.

Now it is very important to realize that no other $i \epsilon$ insertion is possible on the Lorentzian mode. If we would change it on the Lorentzian mode we must change it on the Euclidean segments accordingly. But such a change on the Euclidean segment is not allowed, since it would introduce a singularity in either $\Psi^{\pm}_0(\tau_0, \vec{x}, u)$ or $\Psi^{\pm}_2(\tau_2, \vec{x}, u)$. For instance, if we replace $ + i \epsilon$ by $ - i \epsilon$ on $M_2$, then $\Psi^{\pm}_2(\tau_2, \vec{x}, u)$ is singular at $\tau_2 = \epsilon/2T$, around the point given by $\vec{x}^2 + u^2 = T^2$. After some meditation we conclude, that the $i \epsilon$-insertion in \ref{pos_prop} is the \emph{only} one which moves the singularity everywhere away from the integration contour!

We split the contour-integrated action into 
\begin{equation}
S = -\int^0_{-\infty} \!{d\tau_0 L_E(\Psi_{[0]})} + \int^T_{-T} \!{d\tau_0 L_L(\Psi_{[1]})} - \int^\infty_0 \!{d\tau_2 L_E(\Psi_{[2]})}
\end{equation}
with the Lagrangians
\begin{align}
L_L(\Psi) &= i \sqrt{-g} (\overline{\Psi} \Gamma^{M} D_M \Psi - m \overline{\Psi} \Psi) \\
L_E(\Psi) &= - \sqrt{g} (\overline{\Psi} \Gamma^{M} D_M \Psi - m \overline{\Psi} \Psi)
\label{eq:Lagr}
\end{align}
Next, we require continuity of fields and conjugate momenta (1-point functions) on the gluing surfaces, which corresponds to the continuity of $\Psi_+$ and $\Psi_-$:
\begin{align}
\Psi_{[0] \pm} (\tau_0 = 0, \vec{x},u) &= \Psi_{[1] \pm} (t_1 = -T, \vec{x},u), \\
\Psi_{[1] \pm} (t_1 = T, \vec{x},u) &= \Psi_{[2] \pm} (\tau_2 = 0, \vec{x},u).
\label{eq:match}
\end{align}

Note, that Weyl projections are the same on Lorentzian and on Euclidean segments. Since both of them must be hermitian no additional factors of $i$ are possible.

Next, we are going to show that in order to satisfy these matching conditions no normalizable modes can be added to the (\ref{eq:propagator}). Try to add some normalizable modes
\begin{align}
Y^{\pm}_1(t, \vec{x}, u) &= \frac{1}{(2 \pi)^d} \int \!{d\omega} \int \!{d\vec{k} e^{- i \omega t + i \vec{k} \vec{x}} A^{\pm}_{[1]}(\omega, \vec{k}) u^{\frac{d+1}{2}} J_{m\mp 1/2} (\left|q\right| u) \theta(-q^2)}, \\
Y^{\pm}_0(\tau_0, \vec{x}, u) &= \frac{1}{(2 \pi)^d} \int \!{d\omega} \int \!{d\vec{k} e^{\left| \omega \right| \tau_0 + i \vec{k} \vec{x}} A^{\pm}_{[0]}(\omega, \vec{k}) u^{\frac{d+1}{2}} J_{m\mp 1/2} (\left|q\right| u) \theta(-q^2)}, \\
Y^{\pm}_2(\tau_2, \vec{x}, u) &= \frac{1}{(2 \pi)^d} \int \!{d\omega} \int \!{d\vec{k} e^{ -\left| \omega \right| \tau_2 + i \vec{k} \vec{x}} A^{\pm}_{[2]}(\omega, \vec{k}) u^{\frac{d+1}{2}} J_{m\mp 1/2} (\left|q\right| u) \theta(-q^2)} 
%\label{eq:}
\end{align}
Continuity condition between $M_0$ and $M_1$ is $Y^{\pm}_0(\tau_0 = 0, \vec{x}, u) = Y^{\pm}_1(t = -T, \vec{x}, u)$. Although it does not imply equality of integrands immediately, but note that the modes $u^{d/2} J_l(q u)$ are orthogonal
\begin{equation}
\int_0^{\infty} \!{d u u^{-1} J_l(k u) J_l(k' u)} = c \delta(k - k'),
\label{eq:}
\end{equation}
with $c$ a constant. Thus we can equate  the integrands (up to $\omega \leftrightarrow  - \omega$)
\begin{equation}
A^{\pm}_{[1]} (\omega, \vec{k}) e^{- i \omega T} + A^{\pm}_{[1]} (-\omega, \vec{k}) e^{ i \omega T} = A^{\pm}_{[0]} (|\omega|, \vec{k})
\label{12}
\end{equation} 
As we already know, $A^+$ and $A^-$ coefficients are not independent of each other (\ref{eq:rel}):
\begin{equation}
A^+ = \frac{i \not{q} A^-}{2 m - 1},
%\label{eq:}
\end{equation}
which results in
\begin{equation}
\omega A^{-}_{[1]} (\omega, \vec{k}) e^{- i \omega T} - \omega A^{-}_{[1]} (-\omega, \vec{k}) e^{ i \omega T} = |\omega| A^{-}_{[0]} (|\omega|, \vec{k}).
\label{11}
\end{equation}
Multiplying \ref{12} by $\omega$ and comparing it with \ref{11} we conclude, that $A^{-}_{[1]} (-\omega, \vec{k})$ and hence also $A^{+}_{[1]} (-\omega, \vec{k})$ must vanish for positive $\omega$. Analogously, imposing continuity of fields and momenta on the boundary between $M_1$ and $M_2$ implies also vanishing of $A^{\pm}_{[1]} (\omega, \vec{k})$. Physically it naturally means that only negative frequencies are allowed to the past of the sources, and only positive frequencies - in the future. Thus, there are no normalizable states we can add to the propagator on the Lorentzian piece of the bulk. (\ref{eq:propagator}) is unique! Thus we found the unique modes on the entire manifold. The rest is the same as in the Euclidean case. Matching conditions have produced correct $i \epsilon$ insertions!
\subsection{Other Propagators}
In last subsection we have shown which $i \epsilon$ insertion in $q_\epsilon = \sqrt{-\omega^2 + \vec{k}^2 - i \epsilon}$ yields the Feynman or time-ordered propagator, i.e. gives the correct path in the $\omega$-plane around the poles. From here it is easy to understand which insertions are needed in order to get time-reversed, retarded and advanced propagators. For the reference we write them here.

For the time-reversed propagators we must replace $\epsilon$ by $-\epsilon$ in the propagator, i.e. replace $q_\epsilon = \sqrt{-\omega^2 + \vec{k}^2 - i \epsilon}$ by $q_{-\epsilon} = \sqrt{-\omega^2 + \vec{k}^2 + i \epsilon} = \overline{k_{\epsilon}}$ in (\ref{eq:propagator}) and get 
\begin{align}
&X^{\pm}_{1,\text{time-reversed}}(t, \vec{x}, u) \nonumber \\ &= \frac{1}{(2 \pi)^d} \int _{C}\! d\omega \int \! d\vec{k} e^{- i \omega t + i \vec{k} \vec{x}}\frac{q_{-\epsilon}^{m \mp 1/2}}{2^{m-1 \mp 1/2} \Gamma(m \mp \frac{1}{2})} u^{\frac{d+1}{2}} K_{m \mp 1/2}(q_{-\epsilon}u).
%\label{eq:propagator}
\end{align}

For the retarded propagator the correct pole structure is given by $q_{\text{ret}} = \sqrt{-(\omega + i \epsilon)^2 + \vec{k}^2}$ (both poles are below the real axis) and 
\begin{align}
 &X^{\pm}_{1,\text{ret}}(t, \vec{x}, u) \nonumber \\ & = \frac{1}{(2 \pi)^d} \int _{C}\! d\omega \int \! d\vec{k} e^{- i \omega t + i \vec{k} \vec{x}}\frac{q_{\text{ret}}^{m \mp 1/2}}{2^{m-1 \mp 1/2} \Gamma(m \mp \frac{1}{2})} u^{\frac{d+1}{2}} K_{m \mp 1/2}(q_{\text{ret}}u).
%\label{eq:propagator}
\end{align}

Finally, we get advanced propagator defining $q_{\text{adv}} = \sqrt{-(\omega - i \epsilon)^2 + \vec{k}^2}$ (both poles are above the real axis). And again
\begin{align}
&X^{\pm}_{1,\text{adv}}(t, \vec{x}, u) \nonumber \\ &= \frac{1}{(2 \pi)^d} \int _{C}\! d\omega \int \! d\vec{k} e^{- i \omega t + i \vec{k} \vec{x}}\frac{q_{\text{adv}}^{m \mp 1/2}}{2^{m-1 \mp 1/2} \Gamma(m \mp \frac{1}{2})} u^{\frac{d+1}{2}} K_{m \mp 1/2}(q_{\text{adv}}u).
%\label{eq:propagator}
\end{align}

\subsection{Thermal Contour} \label{thermal}
AdS/CFT correspondence is very important tool for studying the strongly coupled systems at finite temperature (and density). Standard approaches (like lattice gauge theory) at current stage of development cannot produce reliable results for such systems. Hence at the moment AdS/CFT is the best approach for investigating interesting temperature effects of such strongly coupled systems as quark-gluon plasma, superconductors, superfluids, etc.

To introduce finite temperature in QFT one needs to compactify the time direction. In AdS/CFT it can be done in two inequivalent ways. First, one can put a black hole in the bulk and associate the Hawking temperature to the temperature on the field theory side. The spacetime gets curved and there appears a compact direction in the boundary. Second, one can compactify one of the boundary directions by hand, i.e. one can calculate correlation functions in thermal ensemble (and not in the vacuum). In fact, for a given temperature only one of these mechanisms can give a consistent result. There is the so called Hawking-Page transition between these two regimes \cite{Witten:1998zw}.

In the context of real-time holographic renormalization we are not interested in the backgrounds with a horizon, since then there are no modes, but only quasinormal modes, i.e. all the poles are away from the real axis and there is no question about choosing the contour or $i \epsilon$-insertions. 

\begin{figure}
\centering
\includegraphics{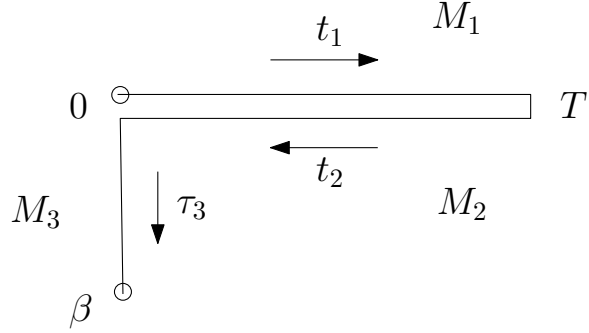}
\caption{Keldysh - Schwinger contour for calculating thermal propagator. Figure is taken from \cite{Leigh:2009eb}.}
\label{thermal-contour}
\end{figure}

Here we calculate a correlation function in a thermal ensemble. To compute a thermal correlator we take the Keldysh-Schwinger contour with the time direction to be compact of period $\beta$ (see figure \ref{thermal-contour}). Fermionic fields must satisfy antiperiodic boundary conditions: $\Psi(0) = - \Psi(-i \beta)$. Denote the segments by $M_1 (t_1 \in \left[0,T\right])$, $M_2 (t_2 \in \left[T, 2 T\right])$ and $M_3 (\tau_3 \in \left[0,\beta \right])$. We place a $\delta$-function source at $t_1=\hat{t}_1$, $\vec{x} = 0$. We make an educated guess and look for a thermal propagator as a linear combination of retarded and advanced propagators
\begin{align}
& \Psi^{\pm}_1(t_1, \vec{x}, u) = \frac{1}{(2 \pi)^d} \frac{u^{\frac{d+1}{2}}}{2^{m-1 \mp 1/2} \Gamma(m \mp \frac{1}{2})} \nonumber \\ & \hspace{1cm} \times \int \! d\omega \int \! d\vec{k} e^{- i \omega (t_1-\hat{t}_1 + i \vec{k} \vec{x})} \big( A(\omega,\vec{k}) q_{\text{ret}}^{m \mp 1/2}  K_{m \mp 1/2}(q_{\text{ret}}u) \nonumber \\ & \hspace{5.4cm} + B(\omega,\vec{k}) q_{\text{adv}}^{m \mp 1/2}  K_{m \mp 1/2}(q_{\text{adv}}u) \big).
%\label{eq:}
\end{align}
with the so far unknown coefficients $A$ and $B$. In order for this to correspond to a $\delta$-function source we must have $A + B = 1$ ($B = -A$ gives a normalizable mode). 

On other segments only normalizable modes are allowed and we make an ansatz for the modes there
\begin{align}
\Psi^{\pm}_2(t_2, \vec{x}, u) &= \frac{1}{(2 \pi)^d} \frac{u^{\frac{d+1}{2}}}{2^{m-1 \mp 1/2} \Gamma(m \mp \frac{1}{2})} \nonumber \\ & \hspace{-1cm} \int \! d\omega \int \! d\vec{k} e^{- i \omega (2T - t_2 - \hat{t}_1) + i \vec{k} \vec{x}} C(\omega, \vec{k}) q^{m \mp 1/2}  J_{m \mp 1/2}(q u) \theta(-q^2)\\
\Psi^{\pm}_3(\tau_3, \vec{x}, u) & = \frac{1}{(2 \pi)^d} \frac{u^{\frac{d+1}{2}}}{2^{m-1 \mp 1/2} \Gamma(m \mp \frac{1}{2})} \nonumber \\ & \hspace{-1cm} \int \! d\omega \int \! d\vec{k} e^{- \omega (\tau_3 - i \hat{t}_1)  + i \vec{k} \vec{x}} D(\omega, \vec{k}) q^{m \mp 1/2}  J_{m \mp 1/2}(q u) \theta(-q^2)
%\label{eq:}
\end{align}
with the to be determined coefficients $C$ and $D$.

The gluing conditions are
\begin{align}
\Psi^{\pm}_1 (t_1 = T) &= \Psi^{\pm}_2 (t_2 = T) \nonumber \\
\Psi^{\pm}_2 (t_2 = 2 T) &= \Psi^{\pm}_3 (\tau_3 = 0) \nonumber \\
\Psi^{\pm}_3 (\tau_3 = \beta) &= - \Psi^{\pm}_1 (t_1 = 0).
\label{matching}
\end{align}
Note an important minus sign in the last equation. It is required by the antiperiodicity of thermal correlators and will give rise to the Fermi statistics as we shall see shortly. 

In what follows we will use the following trick to determine unknown coefficients. We multiply (\ref{matching}) with $e^{- i \vec{k}' \vec{x}} J_{m \mp 1/2} (|q'| u)$ with ${q'}^2 = - {\omega'}^2 + \vec{k}'{}^2$ and integrate over $u$ and $\vec{x}$. We shall make use of the following identities for Bessel functions:
\begin{equation}
\int_0^{\infty} \!{dx x J_n(q x) J_n (q' x)} = \frac{1}{q} \delta(q-q')
%\label{eq:}
\end{equation}
and
\begin{equation}
\int_0^{\infty} \!{dx x J_n(a x) K_n (b x)} = \left( \frac{a}{b}\right)^n \frac{1}{a^2 + b^2}.
%\label{eq:}
\end{equation}

Let us first consider the boundary between $M_1$ and $M_2$. There we have an equality
\begin{align}
&\int \! d\omega \int \! d\vec{k} e^{- i \omega (T-\hat{t}_1) + i \vec{k} \vec{x}} u \big( A(\omega,\vec{k}) q_{\text{ret}}^{m \mp 1/2}  K_{m \mp 1/2}(q_{\text{ret}}u) \nonumber \\ & \hspace{4.2cm}+ B(\omega,\vec{k}) q_{\text{adv}}^{m \mp 1/2}  K_{m \mp 1/2}(q_{\text{adv}}u) \big) \nonumber \\ &= \int \! d\omega \int \! d\vec{k} e^{- i \omega (T-\hat{t}_1) + i \vec{k} \vec{x}} u C(\omega, \vec{k}) q^{m \mp 1/2}  J_{m \mp 1/2}(q u).
%\label{eq:}
\end{align}

The computation for the left-hand side gives
\begin{align}
&\int \! \vec{dx}\frac{d\omega}{2 \pi} \frac{d \vec{k}}{(2 \pi)^d} e^{- i \omega (T-\hat{t}_1) + i(\vec{k} - \vec{k}') \vec{x}} \nonumber \\ & \hspace{3cm} \times \int^{\infty}_0 \! du  \big( A(\omega, \vec{k}) q_{\text{ret}}^{m \mp 1/2} u J_{m \mp 1/2}(|q'| u) K_{m \mp 1/2} (|q_{\text{ret}}| u)  \nonumber \\ & \hspace{4.4cm}+ B(\omega, \vec{k}) q_{\text{adv}}^{m \mp 1/2} u J_{m \mp 1/2}(|q'| u) K_{m \mp 1/2} (|q_{\text{adv}}| u) \big) \nonumber \\
&= \int \!{\frac{d\omega}{2 \pi} e^{-i \omega (T-\hat{t}_1)} \left[\frac{A(\omega, \vec{k}') {|q'|}^{m \mp 1/2}}{-{q'}^2 - (\omega + i \epsilon)^2 + \vec{k}'{}^2 } + \frac{B(\omega, \vec{k}') |q'|^{m \mp 1/2}}{-{q'}^2 - (\omega - i \epsilon)^2 + \vec{k}'{}^2}  \right]} \nonumber \\
&= - |q'|^{m \mp 1/2} \int \!{\frac{d\omega}{2 \pi} e^{-i \omega (T-\hat{t}_1)} \left[\frac{A(\omega, \vec{k}')}{(\omega + i \epsilon)^2 - {\omega'}^2} + \frac{B(\omega, \vec{k}')}{(\omega - i \epsilon)^2 - {\omega'}^2}  \right]} \nonumber \\
&= \frac{i |q'|^{m \mp 1/2}}{2 \omega'} \left[ A(\omega', \vec{k}') e^{-i \omega' (T-\hat{t}_1)} + A(-\omega', \vec{k}') e^{i \omega' (T-\hat{t}_1)}\right],
\label{e:1}
\end{align}
where in the last line we closed the contour in the lower half-plane (picking additional minus sign because of the negative orientation) and thus only the first term has support after the source is switched off (as it should be for retarded propagator). 

The computation on the right-hand side yields
\begin{align}
&\int \!{d \vec{x}\frac{d\omega}{2 \pi} \frac{d \vec{k}}{(2 \pi)^d} e^{-i \omega (T-\hat{t}_1) + i(\vec{k} - \vec{k}') \vec{x}}} C(\omega, \vec{k}) \nonumber \\ & \hspace{0.5cm} \times \int^{\infty}_0 \! {du u J_{m \mp 1/2}(|q| u) J_{m \mp 1/2} (|q'| u) \theta(-q^2)} \nonumber \\
&\hspace{0.5cm}= \int \!{\frac{d\omega}{2 \pi} d \vec{k}  e^{-i \omega (T-\hat{t}_1)} \delta(\vec{k} - \vec{k}') \frac{C(\omega, \vec{k})}{|q'|} \delta( q - q') \theta(\omega^2 - \vec{k}^2)} \nonumber \\
&\hspace{0.5cm}= \int \!{\frac{d\omega}{2 \pi}   e^{-i \omega (T-\hat{t}_1)} C(\omega, \vec{k}') \frac{\delta(\omega - \omega') + \delta(\omega + \omega')}{\omega'} \theta(\omega^2 - \vec{k}'{}^2)}\nonumber \\
&\hspace{0.5cm}= \frac{1}{2 \pi \omega'} \left[ C(\omega', \vec{k}') e^{-i \omega' (T-\hat{t}_1)} + C(-\omega', \vec{k}') e^{i \omega' (T-\hat{t}_1)}\right] \theta({\omega'}^2 - \vec{k}'{}^2).
\label{e:2}
\end{align}
Equating (\ref{e:1}) and (\ref{e:2}) we finally get 
\begin{equation}
C(\omega, \vec{k})  = i \pi |q|^{m \mp 1/2} A(\omega, \vec{k}).
%\label{eq:}
\end{equation}

The matching between $M_3$ and $M_1$ is performed likewise, the only difference being that now the advanced propagator gives non-zero contribution and because of the opposite contour orientation we get additional minus sign which cancels another minus coming from antiperiodicity. Altogether,
\begin{equation}
D(\omega, \vec{k}) e^{- \beta \omega}= i \pi |q|^{m \mp 1/2} B(\omega, \vec{k}).
%\label{eq:}
\end{equation}

Matching between $M_2$ and $M_3$ trivially gives
\begin{equation}
C(\omega, \vec{k}) = D(\omega, \vec{k}).
%\label{eq:}
\end{equation}

Last three equation together with $A+B=1$ give
\begin{align}
A(\omega, \vec{k}) &= \frac{1}{1 + e^{-\beta \omega}}, \\
B(\omega, \vec{k}) &= \frac{1}{1 - e^{\beta \omega}}
\label{c}
\end{align}
and
\begin{align}
&\Psi^{\pm}_1(t_1, \vec{x}, u) = \frac{1}{(2 \pi)^d} \frac{u^{\frac{d+1}{2}}}{2^{m-1 \mp 1/2} \Gamma(m \mp \frac{1}{2})} \nonumber \\ & \hspace{1cm} \times \int \! d\omega \int \! d\vec{k} e^{- i \omega (t_1-\hat{t}_1) + i \vec{k} \vec{x}} \big( \frac{q_{\text{ret}}^{m \mp 1/2}  K_{m \mp 1/2}(q_{\text{ret}}u)}{1 + e^{-\beta \omega}}  \nonumber \\ & \hspace{5.3cm}+ \frac{q_{\text{adv}}^{m \mp 1/2}  K_{m \mp 1/2}(q_{\text{adv}}u)}{1 - e^{\beta \omega}}  \big).
\label{e}
\end{align}

We have derived the well-known formula for a thermal correlator
\begin{equation}
\left\langle T \left( O(x) O(x')\right)\right\rangle = -N(\omega) \Delta_{\text{adv}}(\omega, \vec{k}) + (1+N(\omega)) \Delta_{\text{ret}}(\omega, \vec{k})
%\label{eq:}
\end{equation}
It is very satisfactory that the real-time formalism with all its technical details produces some results which we expect to hold quite generally.

(\ref{e}) can be equivalently rewritten as retarded propagator plus ''thermal bath'' contribution:
\begin{align}
&\Psi^{\pm}_1(t_1, \vec{x}, u) = \frac{1}{(2 \pi)^d} \frac{u^{\frac{d+1}{2}}}{2^{m-1 \mp 1/2} \Gamma(m \mp \frac{1}{2})} \nonumber \\ & \hspace{1cm} \times \int \! d\omega \int \! d\vec{k} e^{- i \omega (t_1-\hat{t}_1) + i \vec{k} \vec{x}} \big( q_{\text{ret}}^{m \mp 1/2}  K_{m \mp 1/2}(q_{\text{ret}}u)  \nonumber \\ & \hspace{3cm}+ \frac{q_{\text{adv}}^{m \mp 1/2}  K_{m \mp 1/2}(q_{\text{adv}}u) - q_{\text{ret}}^{m \mp 1/2}  K_{m \mp 1/2}(q_{\text{ret}}u)}{1 - e^{\beta \omega}}  \big).
%\label{e}
\end{align}

For the source different from the $\delta$-function one should replace 1 by the Fourier transform of the source in the numerators of (\ref{c}).

%\subsection{Infalling boundary conditions at the horizon}

%\section{Real-Time Renormalization} \label{renorm}

%\subsection{On-shell Action and Counterterms}

\begin{comment}
Now we consider half-integer $m$. Again, to determine the divergences we plug in the solution (\ref{log:sol}) into $S_{\text{var}}$ and get
\begin{align}
S_{\text{var}} = \int \! d^d x \frac{1}{\epsilon^d} &\big( \overline{c}_1^+ c_2^- \epsilon^d \ln \epsilon (1 + f_{a^+ d^-}) \nonumber \\ &+\overline{c}_2^+ c_1^- \epsilon^{d+2} \ln \epsilon (1 + f_{d^+ a^-}) \nonumber \\ &+\overline{c}_1^+ c_1^- \epsilon^{d+2m+1} (\ln \epsilon)^2 (1 + f_{a^+ a^-}) \nonumber \\ &+  \overline{c}_2^+ c_2^- \epsilon^{d - 2 m + 1}(1 + f_{d^+ d^-}) \big).
%\label{eq:}
\end{align}
For $m \neq 1/2$ we again use (\ref{relations}) and rewrite $S_{\text{var}}$ as
\end{comment}

%\subsection{Corners} \label{corners}

\chapter{Non-relativistic Holography}
\section{Lifshitz Spacetime and Condensed Matter Physics}

One of the research areas in which gauge/gravity duality is successfully applied is condensed matter physics. For a review and further references see \cite{Hartnoll:2009sz}. Probably, the most important feature of condensed matter systems is that they are not relativistic. This property needs to be reflected in the dual theory, i.e. the local symmetry group of the underlying spacetime must be not Lorentzian, but for example Galilean. More specifically, one speaks about anisotropic spaces, i.e. spaces which are invariant under anisotropic scalings
\begin{align}
 x \rightarrow \lambda x, \;\;\;\;\; t \rightarrow \lambda^z t,
\end{align}
where $z$ is called the dynamical exponent.
 Roughly speaking, there are two families of spacetimes which satisfy this invariance condition: so-called Schroedinger spacetime \cite{Balasubramanian:2008dm, Son:2008ye} and Lifshitz spacetime \cite{Kachru:2008yh, Taylor:2008tg} which has the metric (\ref{metric}). The first has the entire Galilean group as its symmetry, but the last do not admit Galilean boosts and a mass operator.

Fermions on the Schroedinger spacetimes were analyzed in \cite{Leigh:2009ck, Akhavan:2009ns}. Here we will consider Lifshitz spacetimes. Theories which do not admit Galilean boosts or a mass operator (and therefore particle number is not conserved) have a number of condensed matter physics applications, including optimally doped cuprates and non-Fermi liquids near the critical point \cite{2005PhRvB..72b4420G}. We first perform the calculations for the analytically solvable case ($z=2, m=0$) and then analyze the structure of divergences for the general $z$ and $m$.

\section{Euclidean Propagator for Massless Fermions on Lifshitz Spacetime with $z=2$}

Now we consider the case of non-relativistic gauge/gravity duality. We can perform analytical analysis for massless spinor on Lifshitz spacetime with $z=2$. We define the projectors $P_{\pm} = \frac{1}{2}(1 \pm \gamma^{u} \gamma^t)$. Then $\Psi_{\pm} = P_{\pm}\Psi$ satisfy $\gamma^{u} \gamma^t \Psi_{\pm} = \pm \Psi_{\pm}$. We consider the case of Euclidean signature first. The equation of motion becomes
\begin{equation}
\left [ u^2 \partial_u^2 - (d+1) u \partial_u + (-\omega^2 u^4 - (k^2 \mp i \omega) u^2 + (\frac{d}{2} + 1)^2 - \frac{1}{4})\right] \psi^{\pm}(u) = 0
\label{Hermite}
\end{equation}
It is interesting to compare this equation to the equation of motion for a scalar on Lifshitz \cite{Taylor:2008tg}. It reduces to (\ref{Hermite}) if one replaces momentum of the scalar $k^2$ by $(k^2 \mp i \omega)$ and identifies the mass of the scalar $m^2$ with $(\frac{d}{2} + 1)^2 - \frac{1}{4}$. Near the boundary we could make the ansatz $\psi(u) = u^{\lambda} (1 + O(u))$. The characteristic exponents are $\frac{d+3}{2}$ and $\frac{d+1}{2}$.

(\ref{Hermite}) is the Hermite equation and has general solution
\begin{comment}
\begin{equation}
 \psi^{\pm}(u) = u^{\frac{d+1}{2}} e^{-\frac{\omega u^2}{2}} \left[ C_1^{\pm} H_{- 2 \alpha^{\pm}}(\sqrt{\omega} u) + C_2^{\pm} F(\alpha^{\pm};1/2;\omega u^2)\right],
\end{equation}
where $C^{\pm}$ are spinors of definite chirality and $\alpha^+ = \frac{k^2}{4 \omega} + \frac{1-i}{4}$, $\alpha^- = \alpha^+ + i/2$. $H$ is Hermite function (generalization of Hermite polynomial) and $F$ is confluent hypergeometric function. A more appropriate form of the solution is 
\end{comment}
\begin{equation}
 \psi^{\pm}(u) = u^{\frac{d+1}{2}} e^{-\frac{\omega u^2}{2}} \left[ c_1^{\pm} F(\alpha^{\pm};1/2;\omega u^2) + c_2^{\pm} u F(\frac{1}{2} + \alpha^{\pm};3/2;\omega u^2)\right],
 \label{sol}
\end{equation}
where $C^{\pm}$ are spinors of definite chirality, $\alpha^+ = \frac{k^2}{4 \omega} + \frac{1-i}{4}$, $\alpha^- = \alpha^+ + i/2$ and $F$ is confluent hypergeometric function.
The first term has characteristic exponent $\frac{d+1}{2}$ and the second $\frac{d+3}{2}$. Both of them are normalizable, which indicates, that two quantization procedures exist. We interpret the second term (with higher power of $u$) as a source, and the first as a responce (one-point function).

The power series expansions of the confluent hypergeometric function is
\begin{equation}
F(a;b;z) = \sum_{n=0}^{\infty}{\frac{(a)_n z^n}{(b)_n n!}}
\end{equation}
\begin{comment}
\begin{align}
H_{\nu}(z) &= 2^{\nu} \sqrt{\pi} \left( \frac{1}{\Gamma(\frac{1-\nu}{2})} F(-\frac{\nu}{2}; \frac{1}{2};z^2) - \frac{2 z}{\Gamma(-\frac{\nu}{2})} F(-\frac{1-\nu}{2}; \frac{3}{2};z^2) \right) \nonumber \\ &= 2^{\nu} \sqrt{\pi} \left(\frac{1}{\Gamma(\frac{1-\nu}{2})} \sum_{k = 0}^{\infty}{\frac{(-\frac{\nu}{2})_k z^{2k}}{(\frac{1}{2})_k k!}} - \frac{2 z}{\Gamma(-\frac{\nu}{2})} \sum_{k = 0}^{\infty}{\frac{(\frac{1-\nu}{2})_k z^{2k}}{(\frac{3}{2})_k k!}} \right),
\end{align}
\end{comment}
where $(a)_n = \frac{a!}{(a-n)!}$ is the Pochhammer symbol.
It is convenient to rewrite $\psi^{\pm}(u)$ as power series:
\begin{equation}
\psi^{\pm}(u) = e^{-\frac{\omega u^2}{2}} u^{\frac{d+1}{2}} \left[ c_1^{\pm} (1 + s_a^{\pm}(u,k)) + u c_2^{\pm} (1 + s_b^{\pm}(u,k)) \right],
\label{ser}
\end{equation}
where $s_{1,2}$ are series in even powers of $u$ starting with $u^2$.

Now we calculate the on-shell action. As usually for fermions the bulk term vanishes and we have only the boundary term. As we will see immediately we do not need any counterterms in this case and thus
\begin{equation}
 S_{\text{bdy}} = \int \!{d^d x \sqrt{\gamma} \overline{\Psi}_+ \Psi_-}.
\end{equation}
We plug (\ref{ser}) in $S_{\text{bdy}}$ and get 
\begin{align}
S_{\text{bdy}} = \int \! d^d x \frac{1}{\epsilon^{d+1}} e^{-\frac{\omega \epsilon^2}{2}} &(\overline{c}_1^+ c_1^- \epsilon^{d+1} (1 + f_{a^+ a^-}) \nonumber \\&+\overline{c}_1^+ c_2^- \epsilon^{d+2}(1 + f_{a^+ b^-}) \nonumber \\&+\overline{c}_2^+ c_1^- \epsilon^{d+2}(1 + f_{b^+ a^-}) \nonumber \\&+  \overline{c}_2^+ c_2^- \epsilon^{d+3}(1 + f_{b^+ b^-}) ),
\end{align}
where we have defined
\begin{equation}
 f_{a^+ b^-} = s_a^+(\epsilon,k)+ s_b^-(\epsilon,k) + s_a^+(\epsilon,k) s_b^-(\epsilon,k)-
\end{equation}
and similarly for $f_{a^+ a^-}$, $f_{b^+ a^-}$, $f_{b^+ b^-}$, all of which are the power series in $\epsilon^2$ starting with $\epsilon^2$. Now we see that as $\epsilon \rightarrow 0$ only the first term remains finite and all the other terms vanish. Thus the on-shell action is
\begin{equation}
 S = \int \!{d^d x \overline{c}_1^+ c_1^- + O(\epsilon)}.
\end{equation}
Renormalized action is $S_{\text{ren}} = lim_{\epsilon \rightarrow 0} S$. It generates the connected correlators for field theory
\begin{equation}
 e^{-S_{\text{ren}}[c_1^-, \overline{c}_1^-]} = \left\langle  exp \left[ \int \!{d^d x (\overline{c}_1^- O + \overline{O} c_1^-)} \right] \right\rangle 
\end{equation}

As usually, the coefficients $c_{1,2}^{\pm}$ are not independent. Plug (\ref{sol}) into the equation of motion (\ref{eq:Dirac}) and multiply by $\gamma^u$ from the left
\begin{align}
&\gamma^u (i \omega u^2 \gamma^t + i k u \gamma^i + u \gamma^u \partial_u - \frac{d+1}{2} \gamma^u)\psi \nonumber \\
&=(i \omega u^2 \gamma^u \gamma^t + i k u \gamma^u \gamma^i + u \partial_u - \frac{d+1}{2}) (\psi_+ +\psi_-) \nonumber \\
&= (i \omega u^2 + i k u \gamma^u \gamma^i + u \partial_u - \frac{d+1}{2}) \psi_+ + (-i \omega u^2 + i k u \gamma^u \gamma^i + u \partial_u - \frac{d+1}{2})\psi_- \nonumber \\
&= (i k \gamma^u \gamma^i (c_1^+ + c_1^-) + (c_2^+ + c_2^-)) u^{\frac{d+3}{2}} +...
%\label{eq:}
\end{align}
which implies
\begin{align}
c_2^- = - i k \gamma^u \gamma^i c_1^+\\
c_2^+ = - i k \gamma^u \gamma^i c_1^-
\label{rel2}
\end{align}

In order to calculate Euclidean propagator we need to impose regularity of the solution in the bulk. For this we need the asymptotic expansion of confluent hypergeometric functions deep in the bulk
\begin{equation}
F(a;b;z) \propto \frac{\Gamma(b)}{\Gamma(b-a)} (-z)^{-a} (1+O(1/z)) + \frac{\Gamma(b)}{\Gamma(a)} e^z z^{a-b} (1+O(1/z))
%\label{eq:}
\end{equation}
Regularity in the bulk is achieved when
\begin{equation}
c_1^+ = -\frac{1}{2 \sqrt{\omega} }\frac{\Gamma(\alpha^+)}{\Gamma(\alpha^+ + 1/2)} c_2^+
\label{rel1}
\end{equation}
Now we are in position to calculate the renormalized correlator
\begin{equation}
 \left\langle O \overline{O}\right\rangle_{ren} = -\frac{\delta c_1^+}{\delta c_1^-}=G(k) \gamma^t ,
\end{equation}
where $G(k)$ is defined by $c_1^+ = - G(k) \gamma^t c_1^-$. Combine (\ref{rel1}) with (\ref{rel2}) and get 
\begin{align}
\left\langle O \overline{O}\right\rangle_{ren} &= -\frac{1}{2} \frac{\Gamma(\alpha^+)}{\Gamma(\alpha^+ + 1/2)} \frac{k}{\sqrt{\omega}} \gamma^u \gamma^i \nonumber \\ &= -\frac{1}{2} \frac{\Gamma(\frac{k^2}{4\omega} + \frac{1-i}{4})}{\Gamma(\frac{k^2}{4\omega} + \frac{3-i}{4})} \frac{k}{\sqrt{\omega}} \gamma^u \gamma^i.
\end{align}

Note, that the modes do not have any poles on the real axis, thus there are no $i \epsilon$-insertions needed for the real-time propagators.

\section{On the Renormalization for Fermions on Lifshitz Spacetimes}

In this section we derive and evaluate the leading terms of the on-shell action and derive the counterterms for general $z$ and $m$. For simplicity we work in Euclidean signature.

For this purpose it is convenient to rewrite the equation (\ref{eom}) in position space:
\begin{align}
&\Big( u^2 \partial_u^2 - (d+z-1)u \partial_u + \nonumber \\ & \big[ u^{2z} \partial_t^2 +(z-1) u^z \gamma^u \gamma^t \partial_t + u^2 \Box\big] + \big(\frac{d+z}{2}\big)^2 - (m - \frac{1}{2} \gamma^u)^2 \Big) \Psi(u,t,z) =0,
\label{eomx}
\end{align}
where $\Box$ contains derivatives only with respect to spatial coordinates. We can decouple these equations introducing the projector $\tilde{\Pi}_{\pm} = \frac{1}{2} (1 \pm \gamma^u \gamma^i \gamma^t)$. In what follows we will not indicate the quantum number $\pm$ under this operator. In this section we will also use second projector $\Pi_{\pm} = \frac{1}{2} (1 \pm \gamma^u)$. Importantly, $\gamma^u$ commutes with $\gamma^u \gamma^i \gamma^t$, i.e. these two matrices can be diagonalized simultaneously.

(\ref{eomx}) is a second order equation. Near the boundary two different scaling behaviors are possible, with the characteristic exponents determined by
\begin{align}
\Delta_{\pm}^{<} = \frac{1}{2} \left( d+z - |2 m \mp 1| \right), \\
\Delta_{\pm}^{>} = \frac{1}{2} \left( d+z + |2 m \mp 1| \right),
%\label{eq:}
\end{align}
where $\pm$ denotes the eigenvalue with respect to $\gamma^u$. Note that the difference $\Delta_{\pm}^{>} - \Delta_{\pm}^{<} = |2 m \mp 1|$ does not depend on $z$ or $d$. According to the usual holographic prescription, $\Delta_-^< $ is the scaling behavior of the source, whereas the response scales as $\Delta_+^> $. 

Next, we discuss the asymptotic behavior of the fermion near the boundary. For notational simplicity we drop $\pm$ indices. The crucial point to notice is that the equation (\ref{eomx}) contains not only integer powers of $u$, but also powers of $u^z$. Because of this our Ansatz for the asymptotic solution is (existence of the projector $\tilde{\Pi}_{\pm}$ allows us to write such an expansion for each eigenspace of $\gamma^u$ separately)
\begin{align}
 \Psi(u,t,x) &= u^{\Delta^<} \psi(u,t,x) + u^{\Delta^>} \tilde{\psi}(u,t,x) \nonumber \\
& = u^{\Delta^<} \sum_{k,l \in \mathbf{N}} u^{2k + l z}  \psi^{(2k + lz)}(t,x) + u^{\Delta^>} \sum_{k,l \in \mathbf{N}} u^{2k + l z} \tilde{\psi}^{(2k + lz)}(t,x).
\end{align}
We expect that the so far unknown functions $\psi^{(2k + lz)}(t,x)$ are local functions of $\psi^{(0)}(t,x)$.

From the Ansatz it is clear that the logarithmic mode, corresponding to conformal anomaly \cite{Skenderis:2002wp}, will appear when $\Delta_{\pm}^{>} - \Delta_{\pm}^{<} = |2 m \mp 1| = 2k + lz$ for some integer $k$ and $l$. In particular, for even $z$ (as for $z=1$) this mode appears when $m$ is half-integer. For $m=0$ the conformal anomaly appears only for $k=0$, $l=1$ and $z=1$. These conditions are different from the analogous condition for the scalars \cite{Taylor:2008tg}, since the equation of motion for the scalar includes only even powers of $u^z$.

The details of the asymptotic expansion depend on the values of $d,z$ and $m$. We are going to consider a couple of representatives cases.

Let us assume, that $1 < z < 2$. Then the asymptotic expansion begins with
\begin{align}
 \psi (u,t,x) =& \psi^{(0)} (t,x) + u^z \psi^{(z)} (t,x) + u^2 \psi^{(2)} (t,x) + u^{2z} \psi^{(2z)} (t,x) \nonumber \\ & + u^{3z} \psi^{(3z)} (t,x) + ...
\nonumber \\ & + u^{2+z} \psi^{(2+z)} (t,x)+... \nonumber \\
& + u^{4} \psi^{(4)} (t,x) + ...
\label{expansion}
\end{align}
By plugging this Ansatz into the equation of motion (\ref{eomx}) we get the expressions for the first $\psi^{(2k + lz)}(t,x)$ in terms of $\psi^{(0)}(t,x)$,
\begin{align}
&\left ( (\Delta^< + z)(\Delta^< - d) + \big(\frac{d+z}{2}\big)^2 - (m - \frac{1}{2} \gamma^u)^2 \right) \psi^{(z)} + (z-1) \gamma^u \gamma^t \partial_t \psi^{(0)} = 0, \\
&\left ( (\Delta^< + 2)(\Delta^< + 2 - d - z) + \big(\frac{d+z}{2}\big)^2 - (m - \frac{1}{2} \gamma^u)^2 \right) \psi^{(2)} + \Box \psi^{(0)} = 0, \\
&\left ( (\Delta^< + 2 z)(\Delta^< + z - d) - + \big(\frac{d+z}{2}\big)^2 - (m - \frac{1}{2} \gamma^u)^2 \right) \psi^{(2z)} \nonumber \\ & \hspace{6cm}+ (z-1) \gamma^u \gamma^t \partial_t \psi^{(z)} +  \partial_t^2 \psi^{(0)}= 0.
\end{align}

Another representative case is $z=2$. First, for non-half-integer $m$ the expansion becomes
\begin{align}
 \Psi(u,t,x) =& u^{\Delta^<} (\psi^{(0)}(t,x) + u^2 \psi^{(2)}(t,x) + u^4 \psi^{(4)}(t,x) + ...) \nonumber \\
& + u^{\Delta^>} (\tilde{\psi}^{(0)}(t,x) + u^2 \tilde{\psi}^{(2)}(t,x) + ...).
\end{align}
Again, $\psi^{(2)}(t,x)$ and $\psi^{(4)}(t,x)$ are determined by
\begin{align}
&\left ( (\Delta^< + 2)(\Delta^< - d) + \big(\frac{d+z}{2}\big)^2 - (m - \frac{1}{2} \gamma^u)^2 \right) \psi^{(2)} \nonumber \\ & \hspace{5cm}+\gamma^u \gamma^t \partial_t \psi^{(0)} + \Box \psi^{(0)} = 0, \\
&\left ( (\Delta^< + 4)(\Delta^< + 2 - d) + \big(\frac{d+z}{2}\big)^2 - (m - \frac{1}{2} \gamma^u)^2 \right) \psi^{(4)} \nonumber \\ & \hspace{4cm}+ \partial_t^2 \psi^{(0)} + \gamma^u \gamma^t \partial_t \psi^{(2)} + \Box \psi^{(2)} = 0.
\end{align}

For illustration let us consider also the case of half-integer $m$. For definiteness, we set $m = 3/2$. Then the expansion takes the form
\begin{align}
 \Psi_+(u,t,x) =& u^{\Delta^<} (\psi_+^{(0)}(t,x) + u^2 (\psi_+^{(2)}(t,x) + \ln u \; \tilde{\psi}_+^{(2)}(t,x)) + ...).
\end{align}
The coefficient $\psi^{(2)}(t,x)$ cannot be determined by the asymptotic analysis since it corresponds to the response. It must be derived from the solution satisfying certain regularity condition in the bulk. For $\tilde{\psi}^{(2)}(t,x)$ we have
\begin{align}
 \Big(\Delta^< + 1 - d -\big(\frac{d+z}{2}\big)^2 - (m - \frac{1}{2} \gamma^u)^2 \Big)\tilde{\psi}^{(2)}(t,x) + \gamma^u \gamma^t \partial_t \psi^{(0)} + \Box \psi^{(0)} = 0.
\end{align}

Next, we determine the counterterms. The on-shell action is
\begin{align}
 S_{\text{on-shell}} = \int d^{d-1}x dt \sqrt{g_{\text{induced}}} (\overline{\Psi}_+ \Psi_-)_{u = \epsilon}.
\label{sons}
\end{align}
In the case of $ 1 < z < 2$, after plugging in the asymptotic solution, (\ref{sons}) leads to
\begin{align}
 S_{\text{on-shell}} = \int & d^{d-1}x dt \;\;\epsilon^{1 - |m - 1/2| - |m + 1/2|} \nonumber \\ &\times \Big(  \overline{\psi}_+^{(0)} \psi_-^{(0)} + \epsilon^z (\overline{\psi}_+^{(z)} \psi_-^{(0)} + \overline{\psi}_+^{(0)} \psi_-^{(z)}) +  \epsilon^2 (\overline{\psi}_+^{(2)} \psi_-^{(0)} + \overline{\psi}_+^{(0)} \psi_-^{(2)}) \nonumber \\ &+ \epsilon^{2z}(\overline{\psi}_+^{(2 z)} \psi_-^{(0)} + \overline{\psi}_+^{(0)} \psi_-^{(2 z)} +  2\overline{\psi}_+^{(z)} \psi_-^{(z)}) +... \Big).
\label{ons}
\end{align}

For $z=2$ one has a similar structure. The on-shell action is divergent for $m>1/2$. We see once again that for $m=0, z=2$ we do not need any counterterms.

Now, we should express the divergent part of the on-shell action only in terms of the source $\psi_-^{(0)}$. This can be done by plugging the asymptotic expansion (\ref{expansion}) back into the first order equation of motion (\ref{eq:Dirac}). 

In position space
\begin{align}
\psi_+^{(2k + l z)} = \frac{\gamma^i \partial_i \psi_-^{(2k + l z)}}{2 m -1 - (2k +lz)},
\end{align}
where for $z \neq 1$ $i$ stands only for spatial directions.

For the construction of counterterms we will need to invert expansion (\ref{expansion}). To second order
\begin{align}
\psi^{(0)} = \epsilon^{- \Delta^{<}}(\Psi + (z-1) M^{-1} \gamma^u \gamma^t \partial_t \Psi), \\
\psi^{(z)} = - \epsilon^{- \Delta^{<} - z}(z-1) M^{-1} \gamma^u \gamma^t \partial_t \Psi,
\end{align}
where $M = (\Delta^< + z)(\Delta^< - d) + \big(\frac{d+z}{2}\big)^2 - (m - \frac{1}{2} \gamma^u)^2$.
For $1<z\leq2$ and $1/2 < m \leq \frac{z+1}{2}$ only the first term in (\ref{ons}) is divergent and the counterterm action is
\begin{align}
 S_{CT} = \int & d^{d-1}x dt \sqrt{g_{\text{induced}}} \frac{\overline{\Psi}_- \gamma^i \partial_i \Psi_-}{2 m - 1}.
\label{sct}
\end{align}
For the other ranges of parameters the defining principles of finding the counterterm remain the same.

Now we can compute renormalized $1$-point function. For definiteness let us take $z=2$ and $m=1$.
\begin{align}
\left\langle \overline{O} \right\rangle &= -\frac{1}{\sqrt{g_{induced}}}\frac{\delta S_{renormalized}}{\delta \psi_-} \nonumber \\ &=-\lim_{\epsilon\rightarrow 0}\frac{1}{\epsilon^{-\Delta^<_-} \sqrt{g_{induced}}}\frac{\delta S_{renormalized}}{\delta \Psi}_- \nonumber \\ &= -\lim_{\epsilon\rightarrow 0}(\overline{\Psi}_+ - \frac{1}{2 m -1}\partial_i \overline{\Psi}_-\gamma^i) \epsilon^{-\frac{d+3}{2}}\nonumber \\ &=-\overline{\tilde{\psi}}_+^{(d+3)}.
%\label{eq:}
\end{align}
In general for operator of dimension $\Delta = \frac{1}{2}(d+z-2m -1)$ we'll have
\begin{align}
\left\langle \overline{O}_{(\Delta)} \right\rangle = \overline{\tilde{\psi}}_+^{(\frac{1}{2}(d+z+2m -1))} + \text{contributions from counterterms}.
%\label{eq:}
\end{align}

(\ref{sct}) is the main result of this section. It allows to perform holographic renormalization in the cases of asymptotically Lifshitz spaces, i.e. when anisotropic scaling behavior appears in the UV region. It is also necessary to have these counterterms for numerical studies of such systems.

\chapter{Conclusions}

In this thesis we considered one particular piece of holographic dictionary, namely how the real-time correlation functions are encoded in the bulk theory. In particular, we studied, how can one get correct $i \epsilon$-insertions. We generalized the formalism introduced in \cite{Skenderis:2008dh, Skenderis:2008dg} for the case of fermions and illustrated it on easy examples. Real-time holography is a particularly interesting tool, since it allows us to study actual dynamics of physical systems, in particular one can consider response of the system to small perturbations. In recent years it was understood that fermionic fields in strongly coupled systems have particularly interesting behavior and that using the techniques of gauge/gravity duality one can understand many interesting and important features of such systems like the quark-gluon plasma, superconductors, non-Fermi liquids, etc. Fermionic fields in real time is a natural marriage of an interesting object with a useful tool.

In the first two chapters we gave a short introduction into the subject of AdS/CFT correspondence and hopefully a pedagogical review of the techniques of holographic renormalization for Euclidean correlation functions and real-time propagators.

The main new results are concentrated in the third and fourth chapters. There we have derived the equation of motion for fermions on general Lifshitz spacetimes and identified the cases when this equation can be solved analytically. For the case of AdS spacetime we constructed time-ordered, time-reversed, advanced and retarded propagators with the correct $i \epsilon$-insertions. Using the Keldysh-Schwinger contour we also calculated a  propagator on thermal AdS.

In another analytically solvable (and also phenomenologically interesting) case (massless fermions on the Lifshitz spacetime with $z=2$) we calculated the Euclidean 2-point function. Since the mode solutions in this case do not include poles on the real axis we did not need to derive $i \epsilon$-insertions. For the case of general $z$ and $m$ we investigated the asymptotic expansion of the fields and obtained the structure of divergences. We also found the covariant counterterm action which cancels the highest order divergence.

The results obtained here can be used for studying strongly coupled systems which approach AdS (Lifshitz) geometry in the UV region. For example, retarded propagator can be used to calculate different transport coefficients in such systems. For the numerical calculation the divergences of the on-shell action and the structure of counterterms we derived are of great importance.

We would like to mention some directions for the future work. First, we need to construct a Lifshitz black hole to describe a non-relativistic field theory at finite temperature. And secondly, it would be interesting to show for the fermions, that for the retarded propagator the real-time prescription of \cite{Skenderis:2008dh, Skenderis:2008dg} is equivalent to the infalling wave boundary condition at the horizon.

\begin{appendix}

\chapter{Bessel Functions}
Here we collect some mathematical facts concerning different Bessel functions.

Series expansion (definition) of Bessel functions are 
\begin{align}
&J_{n}(x) = \left( \frac{x}{2}\right)^n  \sum^{\infty}_{k=0}\frac{(-1)^k}{k! \Gamma(k+n+1)} \left( \frac{x}{2}\right)^{2k}, \\
& Y_{n}(x) = \frac{2}{\pi}J_n(x)\log\frac{x}{2} - \frac{1}{\pi}\left( \frac{x}{2}\right)^{-n} \sum ^{n-1}_{k=0} \frac{(n-k-1)!}{k!} \left( \frac{x}{2}\right)^{2k} \nonumber  \\
& - \frac{1}{\pi}\left( \frac{x}{2}\right)^{n} \sum^{\infty}_{k=0}\frac{(-1)^k}{k! (k+n)!} [\psi(n+k+1) + \psi(k+1)] \left( \frac{x}{2}\right)^{2k}
\end{align}
where $\psi(x) = \Gamma'(x)/\Gamma(x)$ is the digamma function.

The series expansions of modified Bessel functions are
\begin{align}
& I_{n}(x) = \left( \frac{x}{2}\right)^n  \sum^{\infty}_{k=0}\frac{1}{k! \Gamma(k+n+1)} \left( \frac{x}{2}\right)^{2k}, \\ & K_{n}(x) = (-1)^{n-1} I_n(x)\log\frac{x}{2} - \left( \frac{x}{2}\right)^{-n} \sum ^{n-1}_{k=0} \frac{(-1)^{k} (n-k-1)!}{k!} \left( \frac{x}{2}\right)^{2k} \nonumber  \\
& + \frac{(-1)^n}{2}\left( \frac{x}{2}\right)^{n} \sum^{\infty}_{k=0}\frac{\psi(n+k+1) + \psi(k+1)}{k! (k+n)!}  \left( \frac{x}{2}\right)^{2k}
\end{align}

Note that modified Bessel functions are the ordinary Bessel functions of imaginary argument.

For the large values of the argument $x>>1$ we get asymptotic expansions
\begin{align}
& J_n(x) \approx \sqrt{\frac{2}{\pi x}} \cos (x - \frac{n \pi}{2} - \frac{\pi}{4}), \\
& Y_n(x) \approx \sqrt{\frac{2}{\pi x}} \sin (x - \frac{n \pi}{2} - \frac{\pi}{4}), \\
& I_n(x) \approx \frac{e^x}{\sqrt{2 \pi x}} \Big( 1 + \frac{(1-2n)(1+2n)}{8 x} + ... \Big), \\
& K_n(x) \approx \sqrt{\frac{2}{\pi x}} e^{-x}.
\end{align}

\end{appendix}
%%%%%%%%%%%%%%%%%%%%%%%%%%%%%%%%%%%%%%%%%%%%%%%%%%%%%%%%%%%%%
%% BIBLIOGRAPHY AND OTHER LISTS
%%%%%%%%%%%%%%%%%%%%%%%%%%%%%%%%%%%%%%%%%%%%%%%%%%%%%%%%%%%%%
%% A small distance to the other stuff in the table of contents (toc)
\addtocontents{toc}{\protect\vspace*{\baselineskip}}

%% The Bibliography
%% ==> You need a file 'literature.bib' for this.
%% ==> You need to run BibTeX for this (Project | Properties... | Uses BibTeX)
\addcontentsline{toc}{chapter}{Bibliography} %'Bibliography' into toc
%\nocite{*} %Even non-cited BibTeX-Entries will be shown.
\bibliographystyle{unsrt} %Style of Bibliography: plain / apalike / amsalpha / ...
\bibliography{thesis} %You need a file 'literature.bib' for this.

%% The List of Figures
%\clearpage
%\addcontentsline{toc}{chapter}{List of Figures}
%\listoffigures

%% The List of Tables
%\clearpage
%\addcontentsline{toc}{chapter}{List of Tables}
%\listoftables

%%%%%%%%%%%%%%%%%%%%%%%%%%%%%%%%%%%%%%%%%%%%%%%%%%%%%%%%%%%%%
%% APPENDICES
%%%%%%%%%%%%%%%%%%%%%%%%%%%%%%%%%%%%%%%%%%%%%%%%%%%%%%%%%%%%%

%% ==> Write your text here or include other files.
\newpage
{\Large \textbf{Selbstst\"andigkeitserkl\"arung}}

\vskip 2cm
\setlength{\parindent}{0mm}
Hiermit versichere ich, die vorliegende Arbeit allein und selbstst�ndig und lediglich unter Zuhilfenahme der genannten Quellen angefertigt zu haben.

\vskip 2cm

\par
\begingroup
\leftskip=8cm % ggf. verstellen
\noindent
 \line(1,0){160} \\
Jegors Korovins \\
Juli 2010
\par
\endgroup

\end{document}